\documentclass[prd,singlecolumn,amsmath,nobibnotes,nofootinbib,superscriptaddress,preprintnumbers]{revtex4-1}
\usepackage[margin=2cm]{geometry}
\usepackage{amstext}
\usepackage{amsmath}
\usepackage[pdftex]{graphicx}
\usepackage{hhline}
\usepackage{tabularx}
\usepackage{enumitem}
\usepackage{float}
\usepackage{mathtools}
\usepackage[toc,page]{appendix}
\usepackage{amssymb}
\setlist[itemize]{topsep=0pt,after=\newline}

\begin{document}
\title{Hierarchical Bayesian scheme for measuring the properties of dark energy with \mbox{Strong gravitational lensing}}
\date{\today}

\author{Sungeun Oh}
\email{clapclap.physics@gmail.com}
\affiliation{Department of Physics and Astronomy, York University, Toronto, Ontario, Canada M3J1P3}
~
~
\begin{abstract}
Current sky surveys have been conducted very accurately in order to understand our universe. One of the phenomena survey maps provide is gravitational effect. Albert Einstein (1936) first discussed the possibilities of gravitational lensing effects based on his theory of general relativity, which might give us some information about cosmology. We often categorize the effects by micro, weak, and strong lensing types. This paper focuses on the strong gravitational lensing, especially the double-source plane lens systems. It was recently recognized that the double lensing events can provide us with another method for constraining the cosmological parameters. To get a general idea of 'how many of these lensing events can be found', we set up a toy model, then follow up with several Bayesian models in order to perform statistical analysis with artificially drawn datasets. The results showed that the double lenses are, in fact, promising objects to constrain the cosmological parameters.
\end{abstract}

\maketitle

\section{Introduction}
Gravitational lensing effect was predicted by the idea that light can be bent by gravity. Since the force of gravity is weak, one can observe the effect only in cosmological (reasonably large) scales. English astronomer, Arthur Eddington, first observed the effect during the solar eclipse in 1919. Stars in the constellation Taurus were taken as the source and the sun was assumed to be the lens that bends light. As astronomical techniques advanced, we are now able to look further for gravitational lensing events on galactic scales. A Milkyway-like galaxy with a mass $~10^{12}M_{\odot}$ can strongly bend the path of light originating from another luminous object. This lensing effect can be up to several megaparsecs and cause light notinitially along our line of sight to reach us. If the galaxy (lens) and the luminous object (source) is perfectly aligned along our line of sight, one sees a ring formed by the lensing effect, known as the Einstein Ring. The bigger the size of the source, the thicker the ring will look. A double-lensing effect occurs when two light-emitting sources are aligned with a lens, resulting in two rings surrounding the lens. The double-lensing effect is now under the spotlight after cosmologists realized that it can provide information on dark energy through the ratio of its ring sizes. The research in cosmology is mainly about how we could determine the cosmological parameters that can describe our universe. One of the methods is constraining the parameters by cosmological distances. However, this is quite a difficult task since the cosmological distances depend on all the density parameters $\Omega$'s in addition to the Hubble constant $H_o$. What the double-lensing ratio can do is that it can eliminate the hubble constant so we will be left with only the density parameters. In this paper, we explicate the strong lensing events, and present Hierarchical Bayesian Statistics to constrain dark energy from the double-lensing effect. In order to perform the statistics, we first need to construct a toy model to generate the population of the events. $\Lambda CDM$ is favorably chosen for our model. Galaxies are taken as both the lens and the source for our strong lensing events. For simplicity, we approximate all the lenses and sources as singular isothermal spheres (SIS), i.e. a constant velocity dispersion and mass distribution. The observable quantities from a strong lensing event are, in general, the redshift of each lens and source and the size of the Einstein Ring. We define our lensing event to have a lensed image that gives an arc greater than $\pi/2$, and construct a probability cone, the volume of which represents the expected number of the lensing events for a given lens. We then relate the expected number of events with an approximated number density of sources while also considering the possible magnification bias due to the lensing effect. Finally, we generate an artificial dataset that corresponds to the expected number of events, then test whether Hierarchical Bayesian Statistics can accurately re-evaluate our prior values ($\Lambda CDM$ parameters) just by having the dataset. In this paper, we also provide some background knowledge for the strong lensing effect for those cosmologists who are interested in researching this field.

\section{General Relativity}

\subsection{Gravitational Lensing Effect}
~In general relativity, motions of particles or light can be described by space-time metrics. In case of the presence of an object with its mass M, we use Schwarzschild metric.\\
\begin{align}\label{Schwarzschildmetric}
ds^2=g_{\mu\nu}dX^{\mu}dX^{\nu}=-c^2\left(1-\frac{2GM}{c^2}\right)dt^2+\left(1-\frac{2GM}{c^2}^{-1}\right)dr^2+r^2d\theta^2+r^2\sin^2d\phi^2
\end{align}\\
$ds^2$ is the space-time interval which for the light is 0, and $dX^{\mu}(dX^{\nu})$ is a vector of the differential space-time components. If a metric has some symmetry properties (i. e. if there exists any Lorentz invariant component), one can generate equations of conservation by applying Killing vectors. The equation (\ref{Schwarzschildmetric}) has time-translational invariance and rotational invariance which implies conservation of energy and conservation of angular momentum, respectively.\\
\begin{itemize}
\item \textbf{Time-translational invariance}\\
\\
Killing vector : $T=\left[~\partial_t,0,0,0~\right]~~\rightarrow~~T^{\mu}=\left[~1,0,0,0~\right]~~\rightarrow~~T_{\nu}=g_{\mu\nu}T^{\mu}=\left[\left(1-\dfrac{2GM}{rc^2}\right)c^2,0,0,0~\right]$\\
\\
Conserved energy : $c^2=T_{\nu}\dfrac{dX^{\mu}}{d\tau}=\left(1-\dfrac{2GM}{rc^2}\right) c^2\dfrac{dt}{d\tau}~~\rightarrow~~\dfrac{dt}{d\tau}=\dot{t}=\left(1-\dfrac{2GM}{rc^2}\right)^{-1}$
\item \textbf{Rotational invariance}\\
\\
Killing vector : $R=\left[~0,0,0,~\partial_{\phi}~\right]~~\rightarrow~~R^{\mu}=\left[~0,0,0,1~\right]~~\rightarrow~~R_{\nu}=g_{\mu\nu}R^{\mu}=\left[~0,0,0,r^2~\right]$\\
\\
Conserved angular momentum : $j=R_{\nu}\dfrac{dX^{\mu}}{d\tau}=r^2\dfrac{d\phi}{d\tau}~~\rightarrow~~\dfrac{d\phi}{d\tau}=\dot{\phi}=\dfrac{j}{r^2}$
\end{itemize}
The conserved energy $c^2$ and the conserved angular momentum $j$ are both with respect to the proper time $\tau$.
The equation (\ref{Schwarzschildmetric}) now becomes \\
\begin{align}\label{deflectionintegral}
d\phi=\dfrac{j}{r^2}\dfrac{dr}{\sqrt{c^2-\dfrac{j^2}{r^2}\left(1-\dfrac{2GM}{rc^2}\right)}}
\end{align}\\
We are considering a lens and a source being as far as gigaparsecs, which allows 'thin lens approximation'. This approximation assumes that the bending occurs at $r_m$ from the lens, thus, $dr/d\phi=0$. This condition gives
\begin{align}\label{deflectingpoint}
j=\dfrac{r_mc}{\sqrt{1-\dfrac{2GM}{r_mc^2}}}
\end{align}
Hence, by integrating the equation (\ref{deflectionintegral}) with an assumption $r_m>>2GM/c^2$, we get
\begin{align}\label{phiangle}
\phi_m-\phi_{\infty}=\frac{\pi}{2}+\frac{2GM}{r_mc^2}
\end{align}
\begin{figure}[H]
\centering
\includegraphics[scale=0.4]{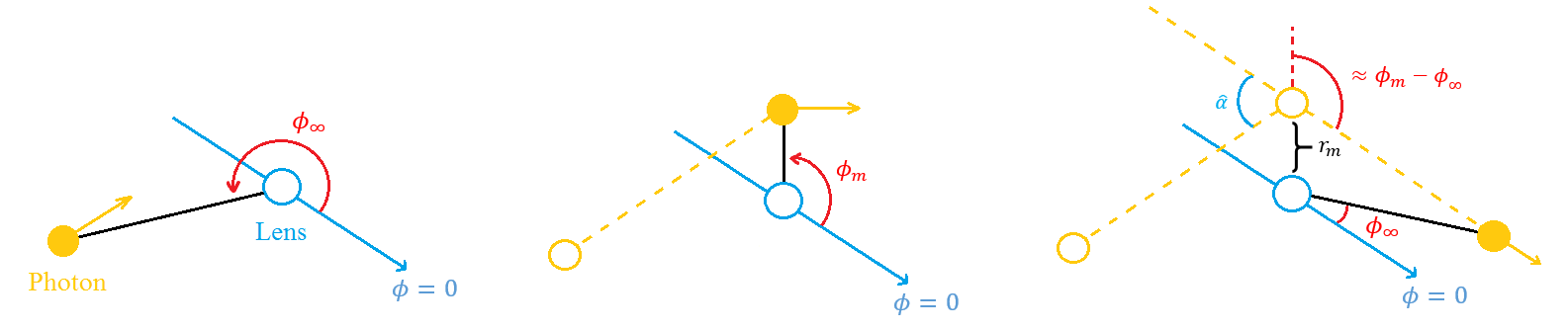}
\footnotesize
\caption{\small First(left) figure shows a photon approaching near the lens, second(middle) shows the photon arriving at $r_m$ and third(right) shows the deflected photon leaving the region.}\label{deflectionangleFigure}
\end{figure}
We denote the deflection angle light  by $\widehat{\alpha}$. FIG \ref{deflectionangleFigure} illustrates how the deflection angle is twice the angle difference subtracted by $180^o$.
\begin{align}\label{deflectionangle}
\widehat{\alpha}\approx2(\phi_m-\phi_{\infty})-180^o=\dfrac{4GM}{r_mc^2}
\end{align}

\subsection{Einstein Ring}
\begin{figure}[H]
\centering
\includegraphics[scale=0.4]{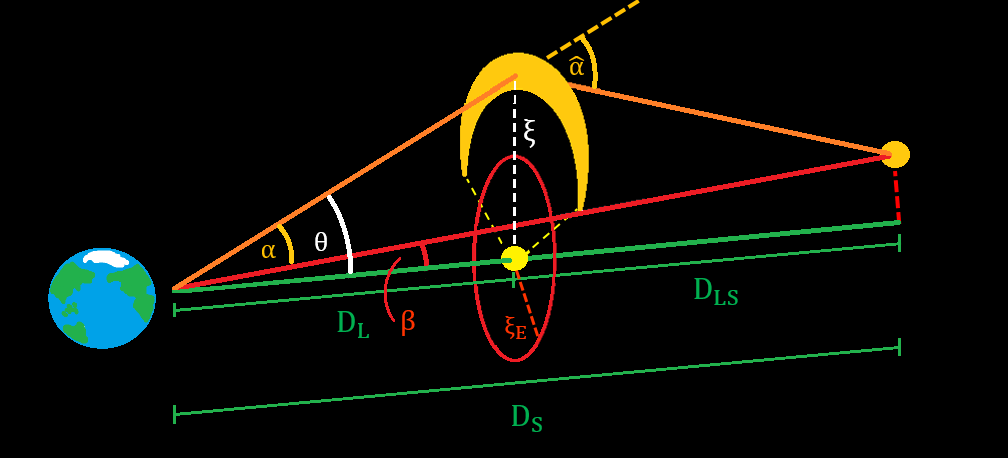}
\footnotesize
\caption{\small Overall illustration of a gravitational lensing event. The green line represents the optical axis with respect to a lens. }\label{LensingIllustrationFigure}
\end{figure}

The formation of Einstein Ring is a geometric consequence of general relativity. We are going to be explaining the physical process with providing an illustration (FIG \ref{LensingIllustrationFigure}). We consider a source located at $D_S$ away from the observer and $\beta$ away from the optical axis, projecting its lensed appearance on the lens plane. The mathematical derivation for the gravitational lensing effects is done in terms of the angular diameter distance. $D_L$ and $D_S$ are the angular diameter distance of a lens and a source, respectively. $D_{LS}$ is the angular diameter distance between the lens and the source. However, this does not really mean $D_{LS}=D_S-D_L$ due to the expansion of the universe. 
A perfect ring is a rare occurance. Fortunately, a slight mis-alignment far less affects the size of the ring compare to the variation in distances between the observer, the lens, and the source. As long as the lensing event occurs, the geometry can still be described in a generic form by the lens equation. \\
\begin{align}\label{lensequation}
\overrightarrow{\theta}=\overrightarrow{\beta}+\overrightarrow{\alpha}~~\longrightarrow~~ \theta\widehat{\theta}=\beta\widehat{\beta}+\alpha\widehat{\theta}
\end{align}\\
Lens and source planes are defined to be the planes perpendicular to the optical axis that centers the lens and source, respectively. We have previously obtained the deflection angle that is derived from general relativity.\\
\begin{align}
\widehat{\alpha}=\frac{4GM(\xi)}{c^2\xi}
\end{align}\\
This describes by how much a light passing by the lens with its mass, $M$, is bent at the distance $\xi$ from the center of the lens. Because of the fact that we are considering SIS objects, if we assume that the bending of light happened outside of the mass shell the total enclosed mass of the lens is $M(\xi)=\pi\sigma^2_v\xi/G$ where $\xi=D_L\tan{\theta}\approx D_L\theta$ by the thin lens approximation. In FIG \ref{LensingIllustrationFigure}, one can see that $D_{LS}\widehat{\alpha}=D_S\alpha$ and yield\\
\begin{align}\label{deflectionangle}
\alpha=\frac{4\pi\sigma^2_v}{c^2}\frac{D_{LS}}{ D_S}
\end{align}\\
In case of the perfect alignment where Einstein Ring appears ($\beta=0$), the equation (\ref{lensequation}) simply becomes $\theta_E\widehat{\theta}=\alpha\widehat{\theta}$ where both vector $\overrightarrow{\theta}$ and $\overrightarrow{\alpha}$ have the same unit vector $\widehat{\theta}$. Therefore, this implies $\theta_E=\alpha$ and we arrive at the final expression for the Einstein Angle.
\\
\begin{align}\label{EinsteinAngle}
\theta_E=\frac{4\pi\sigma^2_v}{c^2}\frac{D_{LS}}{ D_S}
\end{align}\\

\subsection{Calibrating Cosmological Distances}
\begin{figure}[H]
\centering
\includegraphics[scale=0.8]{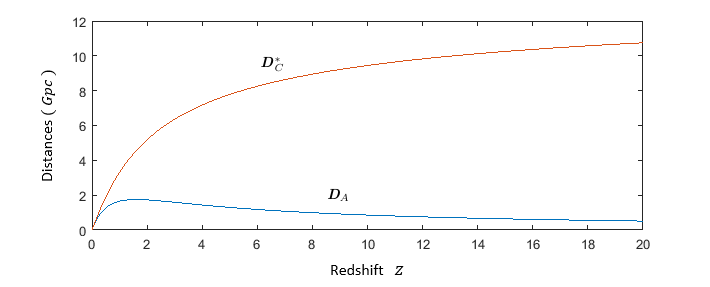}
\footnotesize
\caption{\small Red and blue lines indicate the comoving and the angular diameter distances calibrated using the redshifts, respectively.}\label{calibrationfigure}
\end{figure}

\begin{align}\label{FRWmetric}
ds^2=-c^2dt^2+a^2(t)\left[dr^2+S^2_k(r)\left(d\theta^2+\sin^2\theta d\phi^2\right)\right]
\end{align}\\
In an expanding universe, we have the Friedmann-Robertson-Walker(FRW) metric. $a(t)$ is the scale factor that describes the expansion and $S_k(r)$ is a curvature function that depends on whether the universe is flat, open or closed. Based on the current consensus we assume our universe to be flat and in favor of $\Lambda CDM$ \cite{key11}. For simplicity, we consider a photon that follows radial path (i.e. $ds=d\theta=d\phi=0$). $dr$ is the comoving coordinate that can describe the comoving(unchanged) distance between two different locations. We can derive the comoving distance as follows\\
\begin{align}\label{ComovingDistanceDerivation}
\int^{t_{obs}}_{t_{em}}\frac{dt}{a(t)}=-\int^0_{D^*_C}dr
\end{align}\\
The negative sign on one side is accounting for the fact that light is travelling toward the observer. Our goal is to relate the comoving distance to its redshift. We first re-express the left hand side of the equation(\ref{ComovingDistanceDerivation}) by using the expression for the Hubble parameter corresponding to $\Lambda CDM$\\
\begin{equation}
\setlength{\jot}{12pt}
\begin{aligned}\label{HubbleParameter}
H=\frac{1}{a}\frac{da}{dt}\approx H_o\sqrt{\Omega_{M}a^{-3}+(1-\Omega_{M})a^{-3(1+\omega_{\Lambda})}}\\ \longrightarrow~~\int^{t_{obs}}_{t_{em}}\frac{dt}{a(t)}=\int^{a_{obs}}_{a_{em}}\frac{da}{H_o\sqrt{\Omega_{M}a^{-3}+(1-\Omega_{M})a^{-3(1+\omega_{\Lambda})}}}
\end{aligned}
\end{equation}\\
Matters are treated as dusts $(\omega_M=0)$ so its density parameter is simply $\Omega_Ma^{-3(1+\omega_M)}=\Omega_Ma^{-3}$. The curvature density $\Omega_k$ and the radiation density $\Omega_r$ today are small so the two terms are neglected in this equation. The relation between the scale factor and the redshift is $a(t)=1/(1+z)$, which allows us to finally obtain the expression as follow\\
\begin{align}\label{Comoving}
D^\ast_C=\frac{c}{H_0} \int_{z_i}^{z_j} \frac{~dz}{\sqrt{\Omega_M(1+z)^3+(1-\Omega_{M})(1+z)^{3(1+\omega_{\Lambda})}}}
\end{align} \\
The angular diameter distance corresponds to the proper distance between the observer and the source at the moment when light was emitted from the source.\\
\begin{align}\label{Angular}
D_A=\frac{D^\ast_C}{1+z_j}
\end{align}\\
It is now possible to calibrate the angular diameter distances $D_A$ and the comoving distances $D_C^*$ using redshifts by assuming our universe is flat with the density parameters $\Omega_M=0.3$, $\omega_{\Lambda}=-1$ and Hubble parameter $H_0=70km/s/Mpc$ (FIG \ref{calibrationfigure}).
With respect to these parameters, the angular diameter distance in the graph curves down at the value of redshift about 1.6, so we would want to vary the distance only between redshift 0 and 1.6 throughout the paper.\\
\section{A toy Model For Strong Lens Population}
One would first question how many lens systems are within our field of view that we can observe. According to the Sloan Lens Advanced Camera for Surveys (SLACS), a sample of 131 strong gravitational lenses out of 3 million galaxies have been found so far \cite{key1}. To confirm that this number is close to what we would expect, we introduce a statistical toy model for the strong lens population.\\

\subsection{Parametrization}
In order to simulate a ring image, we first need to parametrize the lens-source system. The parametrization method that is going to be shown here is a simplified version of Dobler \& Keeton (2006) where external shear and convergence will not be considered. In other words, our sources and lens are going to be singular isothermal spheres (SIS). $ (u, v)$ and (r$\cos$$\phi$, r$\sin$$\phi$) are set to be the angular distance coordinates on the source plane and the image place, respectively. 
\\
\begin{align}\label{para1}
\left( \begin{array}{c} u\\v \end{array} \right) =\left( \begin{array}{c} r\cos \phi \\r\sin \phi \end{array} \right) - \left( \begin{array}{c} \theta_E \cos \phi \\ \theta_E\sin \phi \end{array} \right) 
\end{align}\\
The cosine and sine terms are responsible for the tangential and radial mapping, respectively.The radial transformation of the image would not be noticeable by the observer whose observing direction is along the optical axis. Only the tangential transformation would affect observations. Let our finite source have a size $a$, and obtain other parametric equations.
\\
\begin{align}\label{para2}
\left( \begin{array}{c} u\\v \end{array} \right) =\left( \begin{array}{c} u_o\\v_o \end{array} \right) +\left( \begin{array}{c} a \cos \chi \\ a\sin \chi \end{array} \right)
\end{align}\\
~Surface of the source is mapped by $\chi $ on its plane, and $u_o$ and $v_o$ are just possible position offsets that one can add into the equation. The source being farther from the lens is equivalent to the angular size of the source becoming smaller which will make the ring images appear thinner. This means that the source size $a$ can be varied by the source distance $D^\ast_S$. Combining equation (\ref{para1}) and (\ref{para2}) cancels out the angle $\chi$ and gives equations for the inner and outer boundaries of the lensed image, $r_{\pm}(\phi)$. The area of the image can be then computed with these radial boundary equations.
\\
\begin{align}
A=&~2(\theta_E+u_o\cos\phi+v_o\sin\phi) \\
B=&~4(a^2-u_o^2-v_o^2+u_o^2\cos^2\phi+v_o^2\sin^2\phi+2u_ov_o\cos\phi\sin\phi)
\end{align}
\begin{align}
r_\pm (\phi)=\frac{A\pm\sqrt{B}}{2}
\end{align}
\begin{align}
Area=\frac{1}{2}\int_{I} r_+^2(\phi)-r_-^2(\phi)~d\phi
\end{align}\\
Note that the images are projected only where the boundaries are real and positive. When the external shear and convergence are considered, the equation (\ref{para1}) is
\begin{align}
\left( \begin{array}{c} u\\v \end{array} \right) =\left( \begin{array}{c} (1-\kappa-\gamma)~r\cos \phi \\(1-\kappa-\gamma)~r\sin \phi \end{array} \right) - \left( \begin{array}{c} \theta_E \cos \phi \\ \theta_E\sin \phi \end{array} \right) 
\end{align}\\
The external shear is an anisotropic focusing due to tidal gravitational effect, which can distort the image. The convergence is an isotropic focusing due to local matter density in the lens place, which can change the size of the image \cite{key8}.

\subsection{Probability Cone}
~To illustrate the toy model, values for certain parameters were reasonably selected from past survey results. According to the Sloan Lens ACS (SLACS), strong gravitational lensing usually occurs around foreground galaxies between redshifts 0.05 and 0.5 \cite{key2}. We set our foreground galaxy to be at redshift $z_L=0.1$ with the typical velocity dispersion of $200km/s$ \cite{key2}. We consider a simple case where both lens and source are singular isothermal spheres (SIS) with intrinsic diameter of $20 Kpc$. In most papers, the lensing probability is defined as the expected number of lenses within the Einstein cone for agiven source \cite{key4}. It is certainly not a negligible factor, both shear and convergence are assumed to be zero for calculating the galaxy populations just for the sake of simplicity (ie. $\mu=1$). A strong lensing event is normally identified by whether or not a system produces multiple images of its source \cite{key8}. In that case, the boundary of the cone is simply the Einstein radius (Einstein angle in terms of the angular diameter distance). 
\begin{figure}[H]
\centering
\includegraphics[scale=0.5]{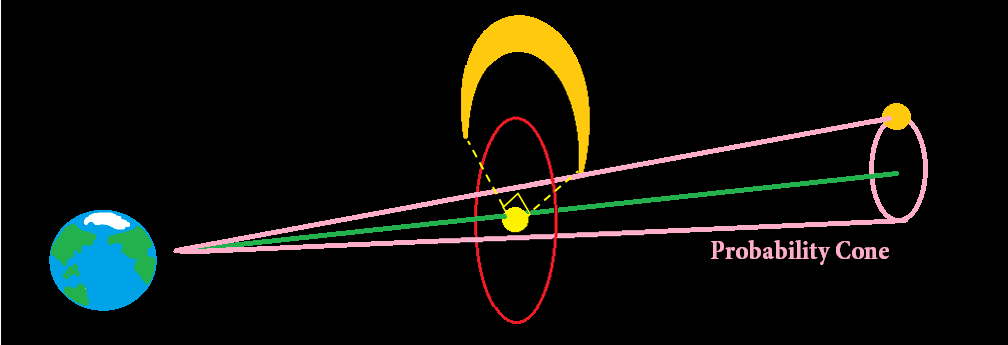}
\footnotesize
\caption{\small As the source moves away from the optical axis, the Einstein ring has reduced its shape to an arc. At a certain point, the arc will become smaller than $\pi$/2 which one could consider it as a boundary of the lensing probability cone.}\label{ConeIllustrationFigure}
\end{figure}
However, one could also identify the lensing event by the images' arc sizes. There are no restrictions on methods for distinguishing Einstein rings from images that do not have full but partial ring shapes (i.e. arcs). Here we will claim that 'it is the ring' if an arc is greater than $\pi/2$. By using this claim and performing a numerical computation, we can generate the equation for the boundary as a function of the source distance, $R_S(D^\ast_S)$. Our cone is illustrated in the FIG \ref{ConeIllustrationFigure}.
\begin{figure}[H]
\centering
\includegraphics[scale=0.8]{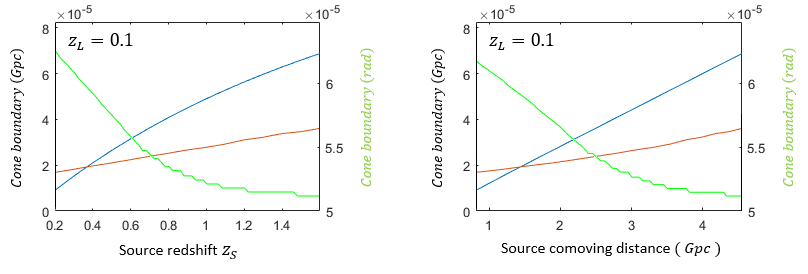}
\footnotesize
\caption{\small The lens is placed at a redshift $z_L=0.1$ from the observer point and the source distance varies in between $z_S=0.12$ and $z_S=1.6$. Einstein radius (blue line), which cosmologists commonly use as the cone boundary, is plotted just for a comparison with our defined boundary (red line). The green line corresponds to our defined boundary in terms of the angular size. Vertical axis represents the boundary of separation angles in units of distances.}\label{ConeRedshift0_1Figure}
\end{figure}
The cross sections enclosed by the image boundaries at source distance $D^\ast_S$ is simply $\pi R_S^2(D^\ast_S)$. By integrating the cross sections over the source distances we obtain a cone-shaped volume which indicates that any source located inside the volume will produce an arc image greater than 90 degrees. In other words, this cone represents the probability density for the presence of an Einstein arc greater than 90 degrees. FIG \ref{ConeRedshift0_1Figure} shows the two differently defined cone boundaries with assuming $(\Omega_M,\omega_{\Lambda})=(0.3,-1)$ and $H_o=68km/s/Mpc$. The size of the probability cone will change with different assumptions for the parameters.
\begin{figure}[H]
\centering
\includegraphics[scale=0.8]{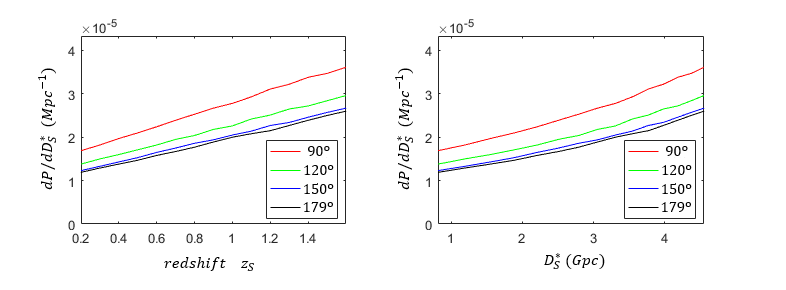}
\footnotesize
\caption{\small The lens is placed at a redshift $z_L=0.1$ from the observer point and the source distance varies in between $z_S=0.12$ and $z_S=1.6$. Vertical axis represents the boundary of separation angles in units of distances.}\label{DifferentRingsConeFigure}
\end{figure}
The size of the probability cone will also change with different minimum sizes for the arc. As shown in FIG \ref{DifferentRingsConeFigure}, the cone will shrink as we are more strict about acceptable arcs (For example, if we would only accept arcs greater than 150 degrees as results of strong lensing events, the probability of detecting such events would be reduced).

If a system contains different values of shear and convergence, it will result in a distortion of lensed images at different strengths, so again, the cone boundary will change. However, for the sake of simplicity we will keep $\gamma=\kappa=0$.
\subsection{Source Population}
\subsubsection{Luminosity Functions}
The population of any type of galaxy is often estimated by the galaxy luminosity function. Certain characteristic numbers that are required to build the function need to be carefully selected from past observation data. There are  three factors that need to be considered for a better picture in a practical sense. \\
\begin{itemize}
\item Galaxies are normally characterized by their flux in a sky survey. This heavily depends on how far they are from us. Due to the technological limitations, we must account for the minimum measurable flux within the survey. 
\item The number density of galaxies are not always the same in every redshift (Ramos et al, 2011) \cite{key3}. 
\item Magnification bias as a result of the gravitational lensing of the sources. 
\end{itemize}
While it would be much simpler to consider a distance-independent number density, we saw that these factors cannot be ignored according to FIG 2.
We start with the Schechter function, which can describe the number densities at certain redshifts, in terms of luminosity.
\\
\begin{align}\label{Schechter}
\Phi(z,L)dL=\phi^{\ast}(z)\left(\frac{L}{L^{\ast}(z)}\right)^{\alpha(z)}e^{-\frac{L}{L^{\ast}(z)}}~d\left(\frac{L}{L^{\ast}(z)}\right)
\end{align}\\
$\phi^{\ast}$ and $L^{\ast}$ refer to the characteristic number density and the characteristic luminosity, respectively, while $\alpha$ is the faint-end slope of the function. We will call these 'evolving variables' since their values tend to change at different redshifts. By integrating the luminosity function, we get the number density function that can give a good approximation for the source population in space. The lower bound of the integral should be the flux limit $F_{obs}=F_o\times\mu$ where $\mu$ is the magnification factor from lensing effect and $F_o$ is the intrinsic flux of the source. We use the luminosity-flux relation $L=4\pi(1+z_S)^4D_S^2F$ and replace the lower bound flux with luminosity. The flux limit $F_o=6\times10^{-20}~Wm^{-2}$ is taken simply in order to compare the result from Dobler (2008) at the end.
\begin{align}\label{ND}
n_S(z_S,L)=\int_{\frac{L_{obs}}{\mu}}^{\infty} \Phi(z_S,L)~dL
\end{align}\\
Our number densities are rather computed in terms of absolute magnitudes $M$ which can be simply obtained from luminosity-magnitude relation 
$M-M^{\ast}=-2.5 \log\left(\frac{L}{L^{\ast}}\right)$.\\
\begin{align}\label{Schechter2a}
\Phi(z,M)dM=0.4\ln10\phi^{\ast}(z)10^{0.4(\alpha(z)+1)(M^{\ast}(z)-M)}e^{-10^{0.4(M^{\ast}(z)-M)}}~dM
\end{align}
\begin{align}\label{Schechter2b}
\Phi(z,M)dM=\phi^{\ast}(z)\left(\frac{M}{M^{\ast}(z)}\right)^{\alpha(z)}e^{-\frac{M}{M^{\ast}(z)}}~d\left(\frac{M}{M^{\ast}(z)}\right)
\end{align}\\
We can use either of the equations, (\ref{Schechter2a}) or (\ref{Schechter2b}), which would give the same results nevertheless.
Data catalog from Canada-France-Hawaii Telescope Legacy Survey (CFHTLS) \cite{key3} are referred to in order to estimate lines of best-fit and obtain expressions for each evolving variables. Almost every detected lensed source is an early-type galaxy (elliptical and lenticular types) according to the Hubble data \cite{key1}, so we will assume our sources to be the early type galaxies as well.\\
\begin{itemize}
\item Characteristic number density : $\phi^{\ast}(z)=(2.17\pm0.25)~10^{-3}~e^{(-6.76\pm1.81)~10^{-1}} ~Mpc^{-3}mag^{-1}$
\item Characteristic luminosity : $M^{\ast}(z)=(-21.65\pm0.01)~e^{(2.30\pm0.32)~10^{-2}~z} ~mag$ 
\item Faint-end slope : $\alpha(z)=-(6.35\pm0.43)~10^{-1}+(4.50\pm4.19)~10^{-2}\log(z)$
\end{itemize}
The provided data catalog has the faint-end slope $\alpha (z)$ greater than -1 which makes the equation (\ref{Schechter2a}) diverge. In this case, one could calculate for the luminosity density instead of just luminosity, then reasonably estimate the number density in terms of luminosity of one specific galaxy (in here, Milky-way's luminosity).
\begin{figure}[H]
\centering
\includegraphics[scale=0.8]{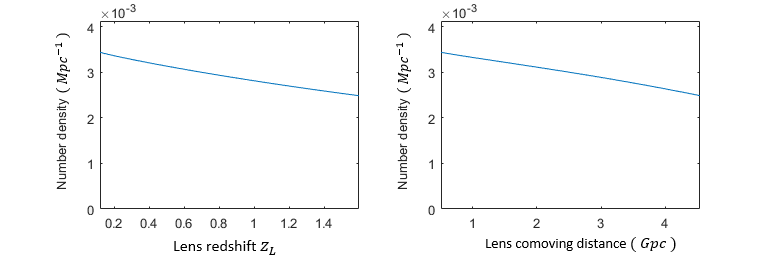}
\footnotesize
\caption{\small Number density of sources at different redshifts(from 0.12 to 1.6). With a reasonably high flux limit, the number density will not change significantly.}
\end{figure}

\subsubsection{Magnification bias}
Magnification bias cannot be ignored when it comes to realistic considerations. Gravitationally lensed images in an observer's view are optically magnified. That can allow the luminosity function to overestimate the number density of a specified type of galaxies. Hence, one needs to modify the luminosity function by including the magnification bias, to calculate the true number densities. Magnification factor from gravitational lensing effect can be derived from the lens equation. For a generic position of the source denoted by $\beta$,  two lensed images will not be located at the Einstein radius but rather at two different positions in the lens plane. 
\begin{figure}[H]
\centering
\includegraphics[scale=0.45]{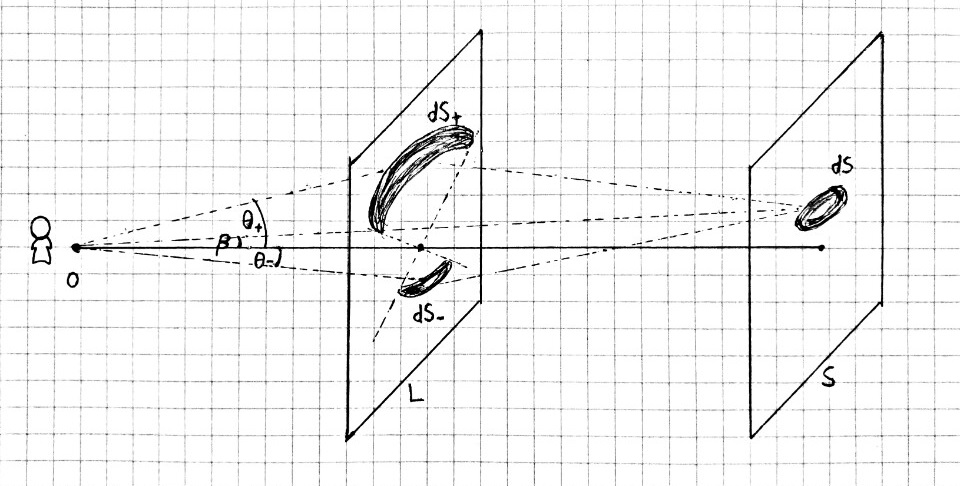}
\footnotesize
\caption{\small Images can be magnified differently with non-perfect alignment of the source.}\label{SchematicFig}
\end{figure}
~
FIG \ref{SchematicFig} shows the case where a source is displayed in two different positions on the image plane (lens plane), subtending solid angles of $d\Omega_{\pm}=dS_{\pm}/D^2_L=\theta_{\pm}d\phi d\theta_{\pm}$. Amplifications of each image with respect to the original source size subtending a solid angle of $d\Omega_0=dS/D^2_S=\beta d\phi d\beta$ is simply $\mu_{\pm}=d\Omega_{\pm}/d\Omega_0=\theta_{\pm}d\theta_{\pm}/\beta d\beta$. Hence, the total magnification is $\mu=|\mu_+|+|\mu_-|$. By using thin approximation to the lens equation $\theta_{\pm}=\beta\pm\theta_{E}$, we obtain the following expression for lensing magnification.  \\
\begin{align}
\mu=\frac{2\theta_{E}}{\beta}
\end{align}
This implies that the magnification is inversely proportional to the angular separation $\beta$ between the source and the optical axis. In other words, magnification factor tends to decrease dramatically as the source is placed farther from the optical axis. 
\begin{figure}[H]
\centering
\includegraphics[scale=0.85]{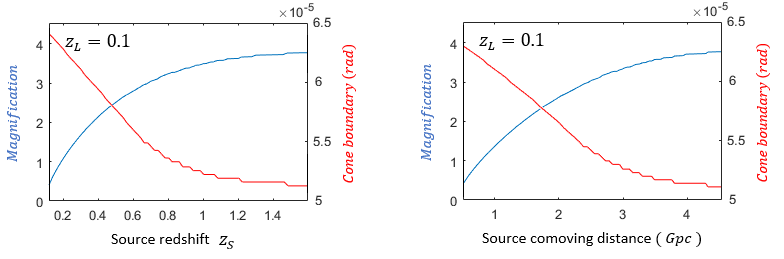}
\footnotesize
\caption{\small Change in magnification at different source positions in the source plane. The lens redshift is set to 0.1 for both figures and the graphs represent magnifications of the cone boundaries (i.e. minimum magnifications, since the cone boundaries are the farthest considered location).}\label{MagCompareFig}
\end{figure}
The equation (\ref{ND}) has the lower boundary which depends on the minimum magnification required to be observable. This minimum magnification also varies with the source distance due to the size of the cone boundary. As shown in the FIG \ref{MagCompareFig}, the minimum magnification increases because its corresponding cone boundary decreases at farther source locations.
  We now have all the ingredients to build a function to calculate for an expected number of the lensing events. The probability density of having a lensing event with one lens is going to be
\begin{align}\label{PD}
\frac{dP_{obs}(D^\ast_L,D^\ast_S)}{dD^\ast_S} = \int_{0}^{r(D^\ast_L,D^\ast_S)_{boundary}} 2\pi r~n_S(D^\ast_S, L/\mu (r))dr
\end{align}
Since the magnification is different at every circumference layer, the probability density first obtained by integrating the number density function over the cone slice (the cross section at $D^\ast_S$). Then the expected number of a lensing event can be obtained by integrating the probability density function over the source distance.
\begin{align}
\frac{dN_{obs,single}}{dD_L}=\frac{d}{dD_L}\int_{D^\ast_L}^{D^\ast_{S,max}} \frac{dP_{obs}(D^\ast_L,D^\ast_S)}{dD^\ast_S}~dD^\ast_S
\end{align}\\
\begin{figure}[H]
\centering
\includegraphics[scale=0.8]{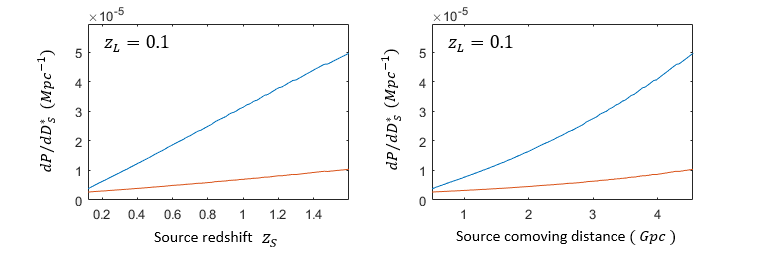}
\footnotesize
\caption{\small The lens is placed at redshift $z_L=0.1$ from the observer point and the source distance varies between $z_S=0.12$ and $z_S=1.6$. Vertical axes represent separation angles which could be converted to arcsecond. Blue line and red line are the probability density with and without magnification biases considered, respectively. }\label{PDcompare}
\end{figure}
 FIG \ref{PDcompare} shows how magnification bias can visibly increase the lensing probability density. Probability of seeing one single lensing event with a lens at $z_L=0.2$ was expected to be $P~\sim~7~\% $ which agrees with the probability from Dobler(2008); $3\%$ without taking the magnification into account. By using the same method for the number density, 31.2 million foreground galaxies are expected to be present within the Hubble field of view of $1400^2$ degrees assuming every galaxy to have the velocity dispersion of $200km/s$ and the luminosity of the Milky-way. The probability of observing a single lensing event with an arc's size greater than $\pi/2$ was estimated to be $0.027$. This implies the detection of one Einstein ring in about 37 foreground galaxies. Comparing with the SLACS survey \cite{key1} which has detected 131 strong gravitational lens samples so far from 3.5 million galaxies of all types, our prediction is far off. It is very unlikely that the immence difference between the toy model and the actual data is due to oversimplification of the toy model. From that, one could easily argue that many more lenses could be found.
\begin{figure}[H]
\centering
\includegraphics[scale=0.7]{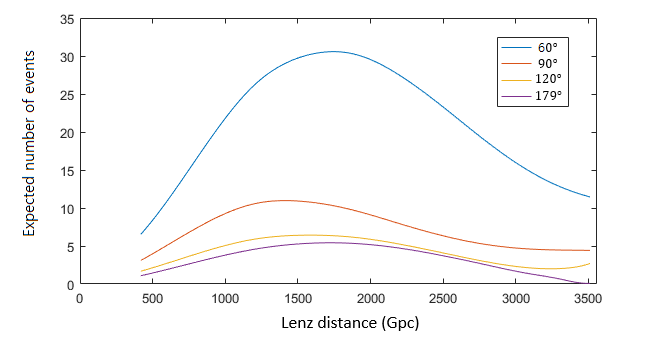}
\footnotesize
\caption{\small We are free to define the strong lensing event by arc sizes other than $\pi/2$. The probability may vary a lot at relatively small arc sizes.}\label{VariousRings}
\end{figure}
As previously mentioned, of course if we have defined the cone boundary differently (from the FIG \ref{DifferentRingsConeFigure}) the expected numbers would change. FIG \ref{VariousRings} shows the probability curves for the different minimum arc size. If we have considered the external shear and convergence, then it can also affect the probability curve.  In FIG \ref{ConvergenceShearFigure}, the locations of the source (colored with teal) are the boundary points for providing a lensed arc image greater than $\pi/2$. In other words, all the images in the figure are showing upper arcs with the size of $\pi/2$. We see that the boundary point is located farther away from the origin with a greater external shear value which implies a change in size of the probability cone.
\begin{figure}[H]
\centering
\includegraphics[scale=0.4]{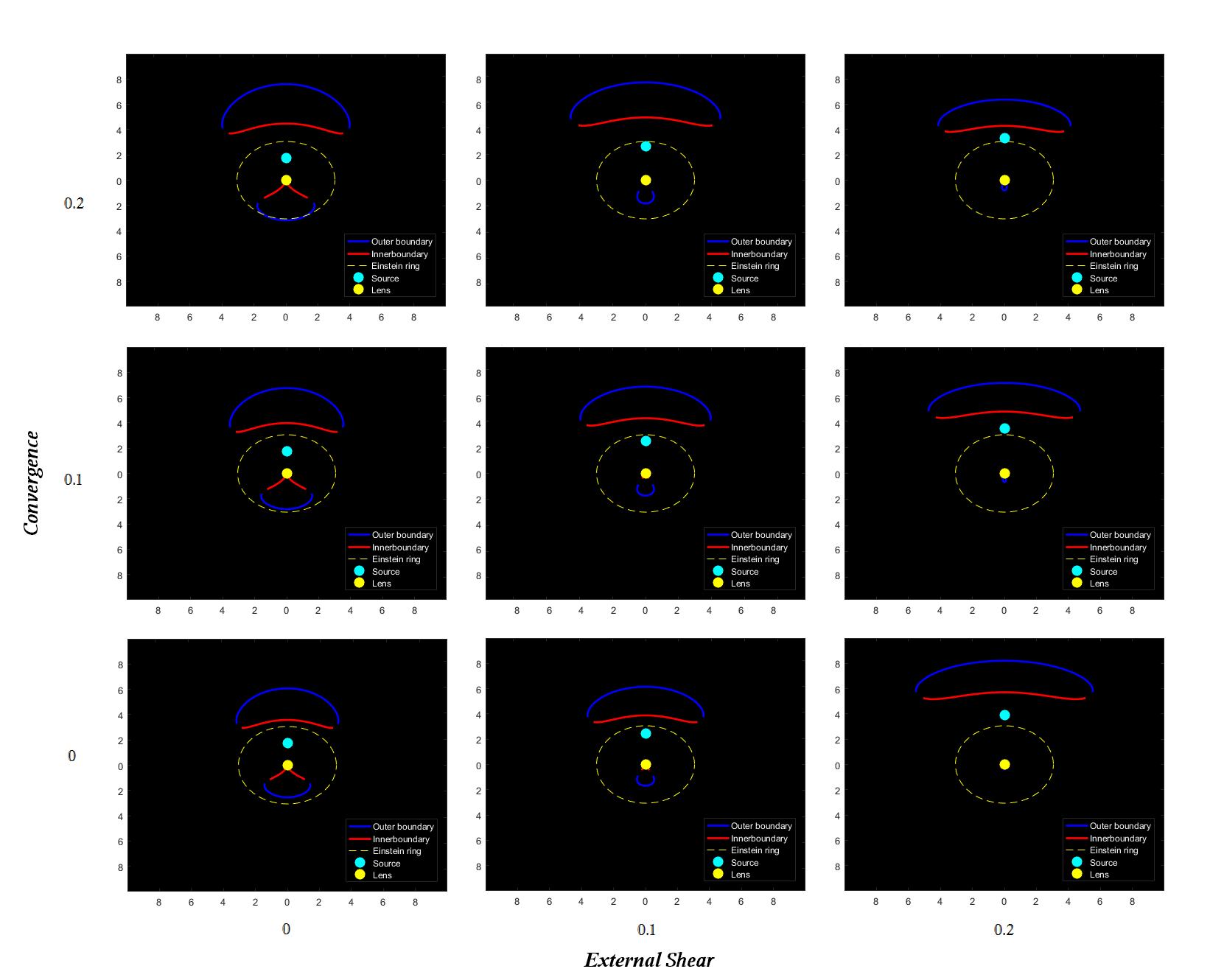}
\footnotesize
\caption{\small Figure chart for different convergence and shear values. The convergence and shear values vary vertically by 0.1 and horizontally by 0.1 in the chart, respectively.}\label{ConvergenceShearFigure}
\end{figure}
~A case where two sources are lensed by the same lens does not require a more complicated method than for the single-lensed. By the rule of multiplication the probability of having the second source at $D^\ast_{S2}$ while the first source is at $D^\ast_{S1}$ is $P(D^\ast_L,D^\ast_{S1},D^\ast_{S2})=P(D^\ast_{S1},D^\ast_L)P(D^\ast_{S2}|D^\ast_{S1},D^\ast_L)$. Positions of the two sources are completely independent from each other, which makes the double lensing probability to simply be a product of two single lensing probabilities (since $P(D^\ast_{S2}|D^\ast_{S1},D^\ast_L)=P(D^\ast_{S2},D^\ast_L))$.\\
\begin{align}\label{Independent}
P(D^\ast_L,D^\ast_{S1},D^\ast_{S2})=P(D^\ast_L,D^\ast_{S1})P(D^\ast_L,D^\ast_{S2})
\end{align}\\
\begin{align}\label{DoublePD}
N_{obs,double}=\int_{0}^{D^\ast_{L,max}}\int_{D^\ast_L}^{D^\ast_{S_1,max}}\int_{D^\ast_L}^{D^\ast_{S_2,max}} \Omega D^{\ast 2}_Ln_L(D^\ast_L,L)\frac{dP_{obs}(D^\ast_L,D^\ast_S)}{dD^\ast_S}\frac{dP_{obs}(D^\ast_L,D^\ast_S)}{dD^\ast_S}~dD^\ast_{S_2}dD^\ast_{S_1}dD^\ast_{L}
\end{align}\\
\\
where $\Omega$ is the solid angle (in radians$^2$) which represents an observing field of view (this is not the density parameter). Considering cases where spreads of lensed images may be greater than their separation angle ratio, the equation (\ref{DoublePD}) could include a step function $\Theta ( > \eta_{min})$ that sets a minimum value for it. As an example, if the full SLACS survey (FOV$=1400^2~deg$) is conducted, about 19,800 double lensing events are expected to occur within the field of view. Since the expected numbers depend on the distances, the numbers should change with different values for the cosmological density parameters. The 3-D plots in FIG \ref{ExpectedNumberPlotFigure} show the distribution of the expected numbers in a logarithmic scale in a 1 degree$^2$ field of view.
\\
\begin{figure}[H]
\centering
\includegraphics[scale=0.35]{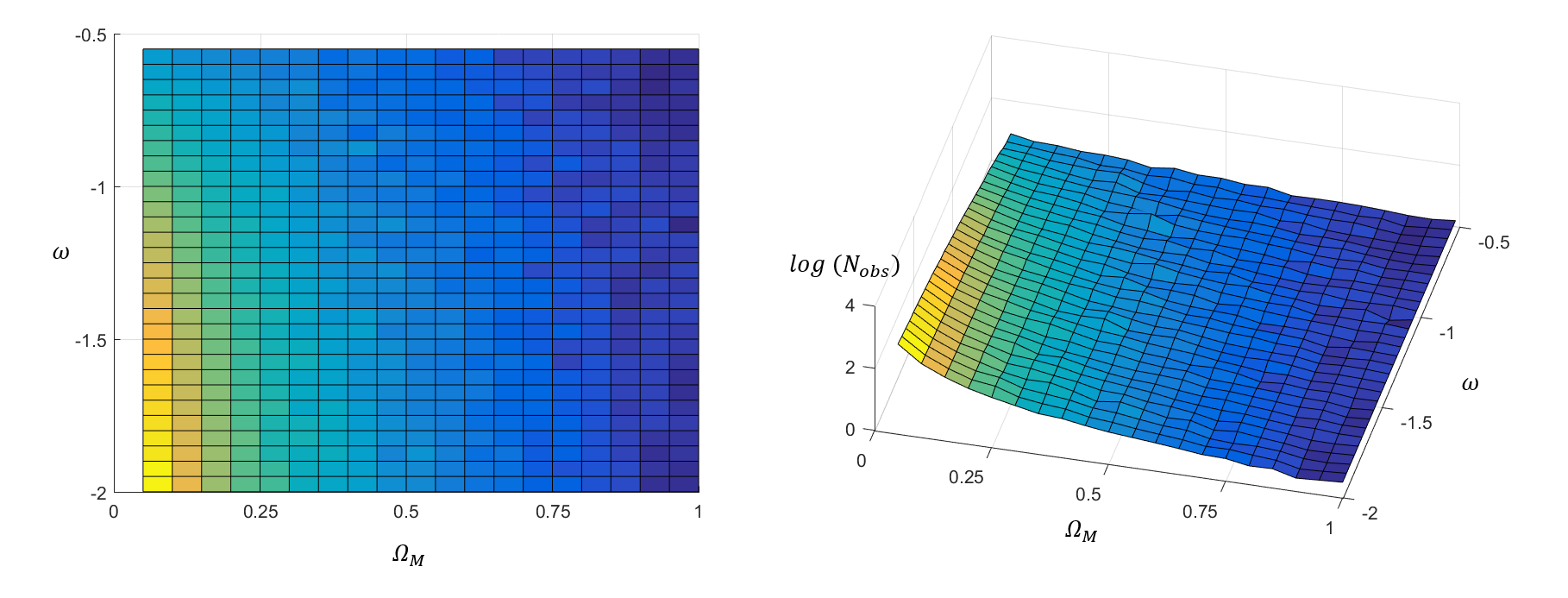}
\footnotesize
\caption{\small Expected numbers of double lensing events from the previous assumptions (all lenses and sources are SIS with equal size, mass and luminosity) plotted in a logarithmic scale. Left and right figures are simply top and side views of the plot, respectively. }\label{ExpectedNumberPlotFigure}
\end{figure}

\section{Constraining Cosmology By Double-Lensed Systems}

\subsection{Lkelihood analysis}
 Formalism of Regular Bayesian Statistics can be adopted in order to understand how well we can constrain the density parameters through a cosmic survey. Bayes' theorem interpret probability as a subjective degree of belief which promotes reasoning with one's hypothesis. Therefore, we can test whether our hypothesis can be confirmed with our set constraint before jumping into an expensive survey project. In this case, we want to constraint the cosmological density parameters, and, in fact, we have already seen that this would be very difficult if our measurements are not done with sufficient precision. This may lead one to perform an extremely costly survey and end up not getting any valuable results. Probability distribution that one's hypothesis is consistent with actual data is called posterior distribution, and is expressed as\\
\begin{align}\label{Bayes' theorem - Expected numbers}
P(X|data)=\frac{P(X)P(data|X)}{P(data)}
\end{align}\\
$P(X_{global})$ is the prior distribution which is going to be one's hypothetical prediction. Throughout the paper, our priors are going to take the ranges of $0\leq \Omega_M \leq 1$ and $-2\leq \omega_{\Lambda}\leq -0.5$ that are 'uniformly distributed (since no restrictions are applied to the parameters)'. $P(data|X)$ is the likelihood of a prior producing the data obtained from the survey, and the term is later divided by $P(data)$ to normalize the posterior result. \\
~
Suppose we want to work out the Bayesian statistics using the expected numbers for each prior. Assuming we can obtain a data for the number of systems observed directly from the survey, the likelihood simply has the form of Poisson distribution \\
\begin{equation}
\setlength{\jot}{10pt}
\begin{aligned}\label{PoissonOnly}
P(\Omega_M,\omega_{\Lambda}|N,f_{sky})~\propto~& P(\Omega_M,\omega_{\Lambda})P(N|\Omega_M,\omega_{\Lambda},f_{sky})\\
=&P(\Omega_M,\omega_{\Lambda})\frac{(f_{sky}\overline{N}(\Omega_M,\omega_{\Lambda}))^{N}}{N!}e^{-(f_{sky}\overline{N}(\Omega_M,\omega_{\Lambda}))}
\end{aligned}
\end{equation}\\
We are looking at one patch of the sky $\Omega=4\pi f_{sky}$ considering no full-sky surveys have yet been done.\\
\begin{figure}[H]
\centering
\includegraphics[scale=0.55]{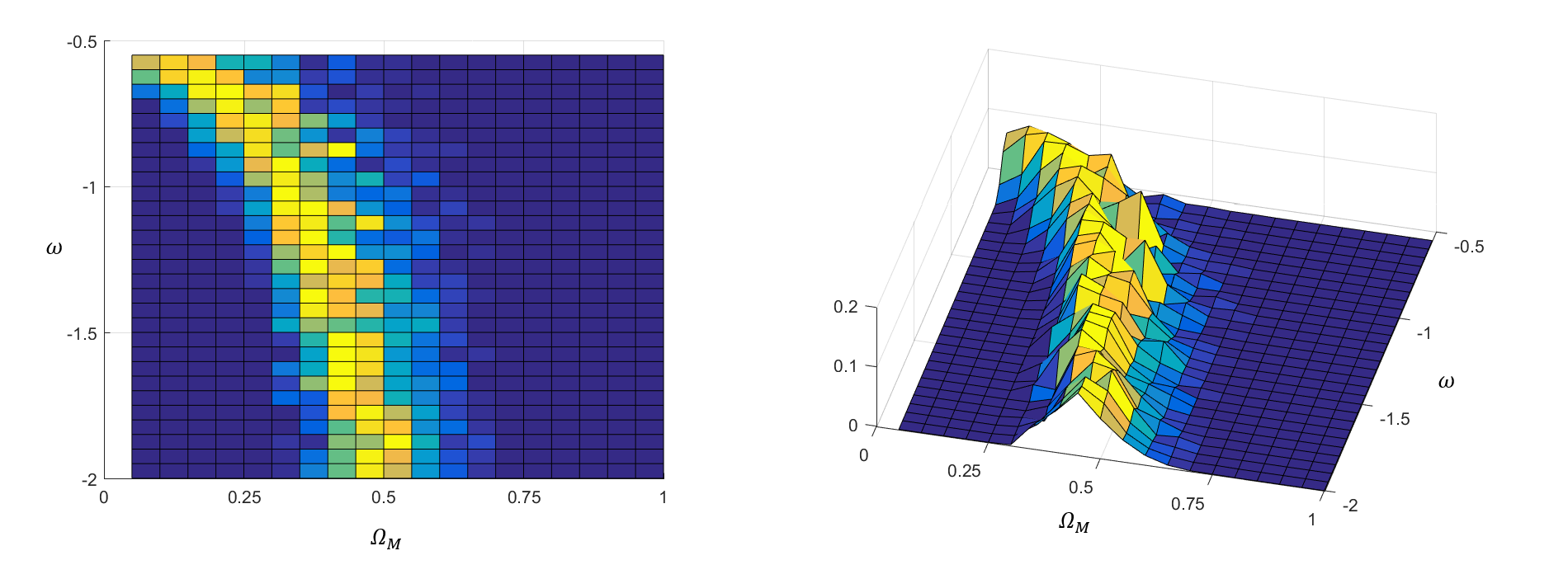}
\footnotesize
\caption{\small In a 1 degree$^2$ field of view, we expect to have 14 systems for $(\Omega_M, \omega_{\Lambda})$=$(0.3, -1)$. Thus, we assumed that we have detected 14 systems through the survey and plotted its posterior distribution. Each prior takes one grid of the plot, and expects different numbers of detections.}\label{PoissonOnlyFigure}
\end{figure}
As shown in Fig \ref{PoissonOnlyFigure}, the posterior showed some degeneracies due to the fact that some priors expect same numbers of detections. The overall distribution could be softened by a better and more accurate calculation for the expected numbers.

\subsection{Likelihood analysis - Angle ratios}
The real reason for investigating strongly double-lensed systems is the unique information they can provide. It was recognized that a double lensing system can be very useful for constraining the cosmology (Collett et al. 2012) \cite{key2}. Say, we see a double ring with radius of each ring denoted by $\theta_{E,S_1}$ and $\theta_{E,S_2}$, and know their redshifts. Not only the ratio of the two can get rid of the $4GM/c^2$ but also the hubble constants $H_0$ which are embedded in each angular diameter distances.
\\
\begin{align}\label{angleratio}
\eta=\frac{\theta_{E,S_1}}{\theta_{E,S_2}}=\frac{D_{LS_1}D_{S_2}}{D_{LS_2}D_{S_1}}=\frac{\int_{z_L}^{z_1} \frac{~dz}{\sqrt{\Omega_M(1+z)^3+(1-\Omega_M)(1+z)^{3(1+\omega_{\Lambda})}}}\int_{0}^{z_2} \frac{~dz}{\sqrt{\Omega_M(1+z)^3+(1-\Omega_M)(1+z)^{3(1+\omega_{\Lambda})}}}}{\int_{z_L}^{z_2} \frac{~dz}{\sqrt{\Omega_M(1+z)^3+(1-\Omega_M)(1+z)^{3(1+\omega_{\Lambda})}}}\int_{0}^{z_1} \frac{~dz}{\sqrt{\Omega_M(1+z)^3+(1-\Omega_M)(1+z)^{3(1+\omega_{\Lambda})}}}}
\end{align}\\
This means that knowing the separation ratio of two rings which are lensed by the same object and the redshifts of each source and the lens will allow one to constrain the density parameters. To explain this in more details, let us assume that $\Omega_M=0.3$, $\omega_{\Lambda}=-1$. Some combinations of redshifts for the lens and sources are randomly drawn. Then the separation ratios are calculated using the equation (\ref{angleratio}).\\
\begin{table}[H]
\centering
\begin{tabular}{ccccc}
\hline
Designation & $z_L$ & $z_{S1}$ & $z_{S2}$ & Separation Ratio\\
\hline
\hline
A & 0.1 & 1.0 & 1.2 & 1.008816662\\
B & 0.1 & 0.8 & 1.2 & 1.022640428\\
C & 0.1 & 1.1 & 1.5 & 1.012571466\\
D & 0.1 & 1.3 & 1.5 & 1.005235356\\
E & 0.1 & 0.7 & 1.3 & 1.036383080\\
F & 0.1 & 0.9 & 1.4 & 1.021102279\\
G & 0.3 & 0.9 & 1.2 & 1.058881102\\
H & 0.5 & 1.1 & 1.3 & 1.060646452\\
I & 0.2 & 0.9 & 1.4 & 1.047885997\\
J & 0.3 & 1.1 & 1.6 & 1.054728296\\
\hline
\end{tabular}\\
\footnotesize
\caption{\small 10 designated double-lensed systems}
\end{table}
~We have previously determined that about 19,800 double-lensed systems will be present in the $1400^2$ degrees Hubble's field of view. Let us reduce the number to 10 and pretend that we have found all the 10 systems from observations and don't know the values for the density parameters. Again, using the equation (\ref{angleratio}), each system will provide us a single line of constraint that represents possible pairs for the density parameters. Putting the constraint lines obtained from the 10 systems on a single plot will give us a good approximation of $\omega_{\Lambda}$ and $\Omega_M$.\\
\begin{figure}[H]
\centering
\includegraphics[scale=0.8]{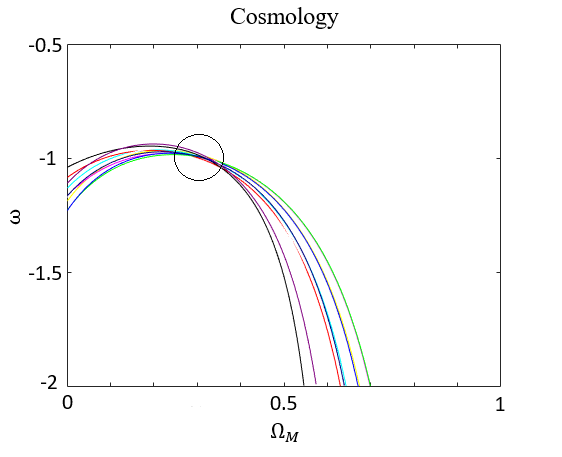}
\footnotesize
\caption{\small Constraint lines from the ten double-lensed systems are overlapped inside the circled region, which presumably indicate the location of the parameter values that we are seeking.}\label{Cosmology}
\end{figure}
~Clearly, we cannot completely constrain the cosmology to thin lines like in FIG \ref{Cosmology} due to limitations in accuracy of measurements. The ''Eye of Horus'' (HSC J142994-005322) which is a double source plane lens system recently discovered had an outer ring with knots offset by $\Delta z=0.002$ which may have simply arose from the uncertainties of the measurements \cite{key7}. To see how this can affect the constraints, we generate several realizations from a gaussian distrubution and see the variations in constraining $\omega_{\Lambda}$ and $\Omega_M$.
\newpage
\begin{figure}[H]
\centering
\includegraphics[scale=0.4]{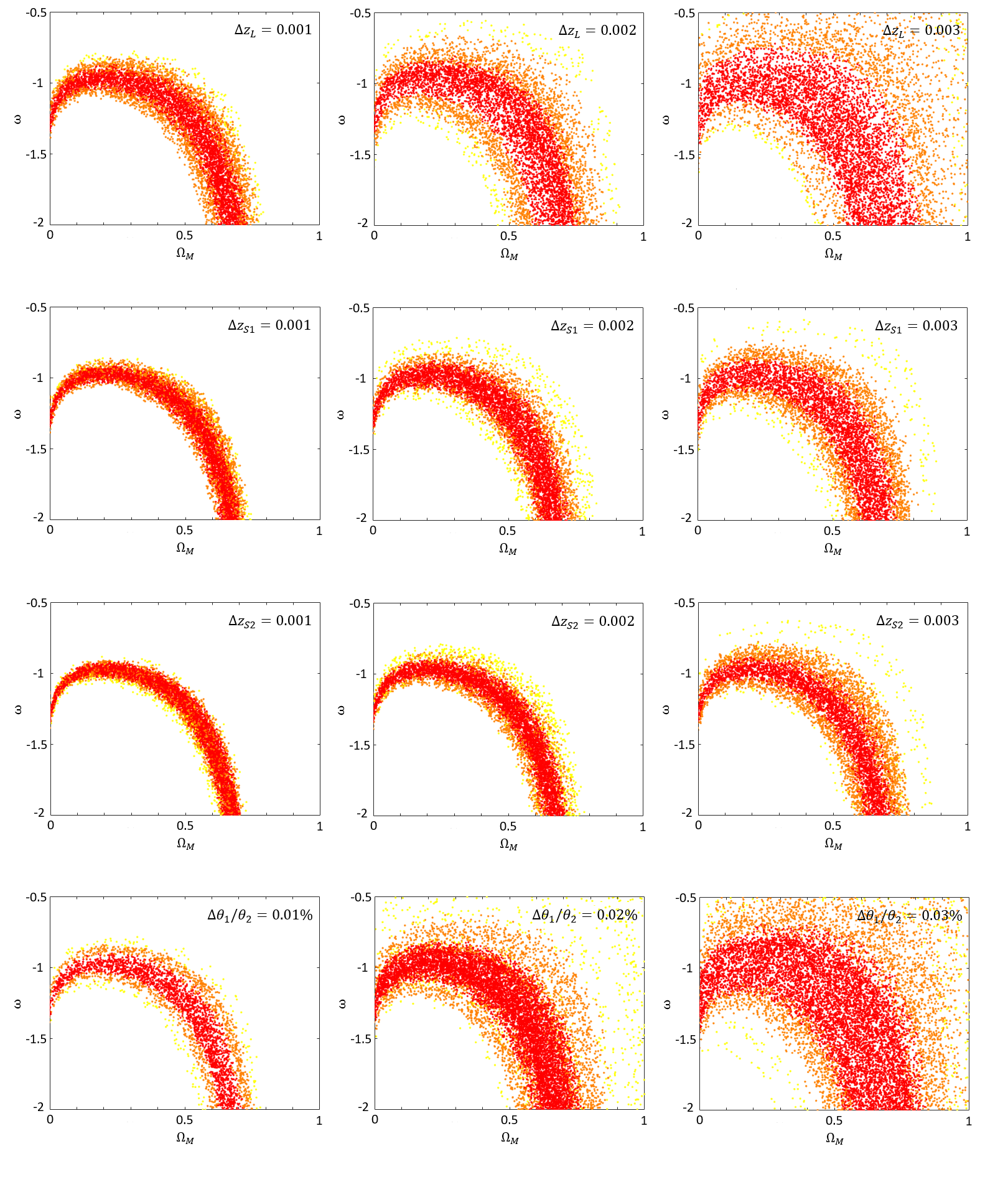}
\footnotesize
\caption{\small Constraint plot for the system 'A' which has $z_L=0.1, z_{S1}=1.0, z_{S2}=1.2$ and $\theta_1/\theta_2=1.009$}\label{LensRealizations}
\end{figure}
FIG \ref{LensRealizations} clearly shows that high uncertainty in the angle ratio of source images and the redshift of a lens poorly constrains the density parameters $\omega_{\Lambda}$ and $\Omega_M$. 500 realizations are drawn for each system and their probability density distributions were combined to show the predicted region of the actual parameter pair. \\
Similarly to the equation (\ref{Independent}), the predicted region is simply going to be the product of all ten systems.\\
\begin{align}\label{IndependentRatios}
P(A\cap B \cdot\cdot\cdot \cap J)=P(A)P(B)\cdot\cdot\cdot P(J)
\end{align}\\
Therefore mathematical expression in the form of Regular Bayesian Statistics is going to be\\
\begin{equation}
\setlength{\jot}{10pt}
\begin{aligned}\label{AngleRatioBayesian}
P(\Omega_M,\omega_{\Lambda}|\eta_1,\eta_2,\cdots,\eta_{10})=&P(\Omega_M,\omega_{\Lambda}|\eta_1)P(\Omega_M,\omega_{\Lambda}|\eta_2)\cdots P(\Omega_M,\omega_{\Lambda}|\eta_{10})\\ 
\propto &~P(\eta_1|\Omega_M,\omega_{\Lambda})P(\eta_2|\Omega_M,\omega_{\Lambda})\cdots P(\eta_{10}|\Omega_M,\omega_{\Lambda})
\end{aligned}
\end{equation}
\\
where each likelihood is going to have the form
\\
\begin{equation}
\setlength{\jot}{10pt}
\begin{aligned}\label{AngleRatioLikelihood}
P(\eta|\Omega_M,\omega_{\Lambda})=\frac{1}{\sqrt{2\pi}\Delta\eta}e^{-\dfrac{{\left(\eta-\eta'(z_L,z_{S1},z_{S2},\Omega_M,\omega_{\Lambda})\right)^2}}{2\Delta\eta}}
\end{aligned}
\end{equation}\\
As shown in FIG \ref{TenCombined}, product of the probability density functions for the ten systems shows the highest probability amplitude at $(\Omega_M, \omega_{\Lambda})=( 0.3, -1 )$ which is precisely the value pair that we have chosen. Unrealistically small values for the deviations of redshifts and angle ratios are taken in here ($\Delta z=0.001$ and $\Delta \eta=0.0001$, which are beyond our current technology precision), so we would actually need a lot more systems instead for an actual survey.
\begin{figure}[H]
\centering
\includegraphics[scale=0.2]{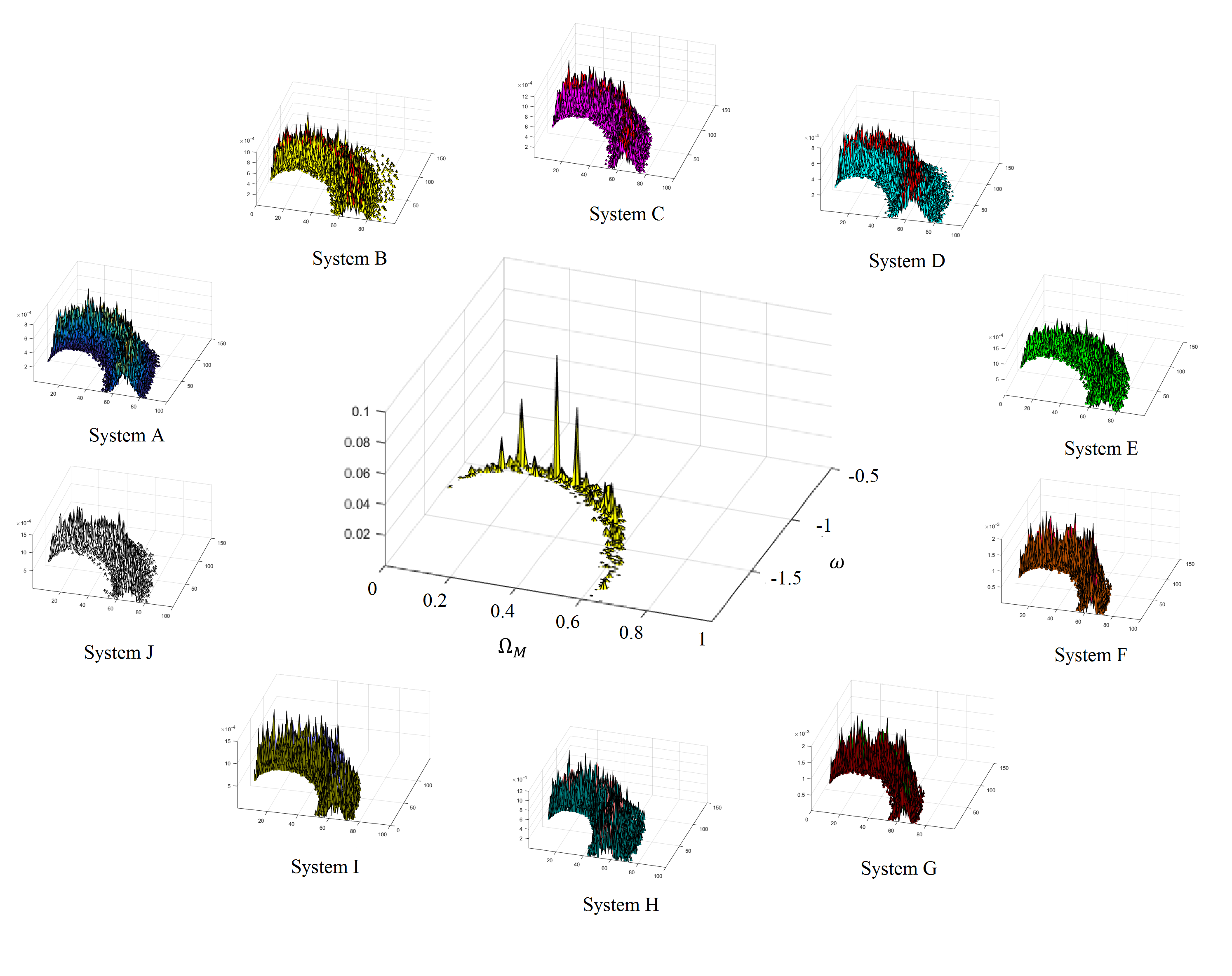}
\footnotesize
\caption{\small The posterior distributions for the System A$\sim$J are multiplied to produce the combined distribution (center plot). }\label{TenCombined}
\end{figure}

\subsection{Hierarchical Bayesian Statistics - Expected numbers}
A double-lensed system provides redshifts of the lens and sources along with the angle ratio between the two ring images. A theoretically, driven value of the angle ratio simply depends on the redshifts, and has no clear restrictions on where the lens and sources should be. There may be more than one set of redshift combination that would produce the same angle ratio (i.e. degeneracy). The fact that there can also be degeneracies in $(\Omega_M,\omega_{\Lambda})$ pair can lead to the assumption that the redshift prior is a flat distribution. As we proceed to a sky survey the expected number of lensing events which depend on the cosmological density parameters will also provides some cosmological constraints. This time, the Regular Bayesian Statistics cannot be directly applied to this quantity. The reason is for that is the expected number varies with the cosmological distances. In other words, the individual system has its own weights that affect the total expected number of systems. The weight comes from the probability cone that was previously introduced. Thus, we need to add local priors that would describe local regions, and make a connection betwen global prior and the data.\\
\begin{align}\label{HierarchicalBayesian}
P(X_{global}|data,f_{sky})=\frac{P(X_{global},f_{sky})\displaystyle\int P(data|X_{local}) P(X_{local}|X_{global},f_{sky})~dX_{local}}{P(data|f_{sky})}
\end{align}\\
 The formalism is going to be analogous to M. Johnson (2013) where cosmic bubbles were the targets, but in our case they are the double ring systems. We are still considering SIS models only, so the probablity density distribution for the double lensing events that we have derived previously is going to be applied here. To apply the Hierarchical Bayesian Statistics, we first introduce the required model parameters. \\
\begin{itemize}
\item Global parameters, $X_{global}$ : These are the ones that are valid in the cosmology as general. The matter density $\Omega_M$ and the dark energy state $\omega_{\Lambda}$ clearly belong to this category. Yet, we have absolutely no information at all about the density parameters, so we may consider a uniform prior distribution described with a window function.
\item Local parameters, $X_{local}$ : We are considering a very simple case where every lens and source have identical properties. In our case, the total number of expected double lens systems  is very much dependent on the density parameters, which implies we cannot use $\overline{N}(\Omega_M,\omega_{\Lambda})$ as another global parameter along with the two density parameters (To obtain a posterior result in terms of the expected number, one has to give up on the other two parameters instead). Each double lens system will have their lenses and sources at different redshifts. Hence, it is better to have a likelihood that describes locally in order to consider the weights. This can be done by defining our local parameters to be the ones that provide spacial information on each individual system; $z_L, z_{S1}$ and $z_{S2}$ which are the redshifts of the lens and the two sources, respectively. Lensing event occurs randomly, so one will not always get the exact same number of events \cite{key10}. One can consider that by simply applying Poisson distribution (which becomes Gaussian distribution at large means) to the expected number. From this moment on, we use $N$=$N(\Omega_M,\omega_{\Lambda},X_{local})$ and $\overline{N}$=$\overline{N}(\Omega_M,\omega_{\Lambda},X_{local})$ as shorthand notations.
\end{itemize}
\begin{equation}
\setlength{\jot}{20pt}
\begin{aligned}\label{BayesianFormalism}
P(\overline{N}|data,f_{sky})~\propto&~P(data|\overline{N},f_{sky})P(\overline{N}|\Omega_M,\omega_{\Lambda})P(\Omega_M,\omega_{\Lambda})\\
=&P(N|\overline{N},f_{sky})\times \left[P(data|N)P(N|\Omega_M,\omega_{\Lambda})P(\Omega_M,\omega_{\Lambda})\right] \\
=&\sum_{N=0}^{\infty}\frac{(f_{sky}\overline{N})^{N}}{N!}e^{-(f_{sky}\overline{N})}\times \left[P(data|N)P(N|\Omega_M,\omega_{\Lambda})P(\Omega_M,\omega_{\Lambda})\right]
\end{aligned}
\end{equation}\\
In order to have a posterior on the global parameter, the enclosed term in the equation (\ref{BayesianFormalism}) must be marginalized over the local parameters. Again, the posterior is going to be directly proportional to the likelihood since the global prior $P(\Omega_M,\omega_{\Lambda})$ has a uniform distribution. 
\begin{align}\label{BayesianFormalismSimplified}
P(\Omega_M,\omega_{\Lambda}|data,f_{sky})~\propto~\sum_{N=0}^{\infty}\frac{(f_{sky}\overline{N})^{N}}{N!}e^{-(f_{sky}\overline{N})} \int P(data|N)P(N|\Omega_M,\omega_{\Lambda})~dX_{local}
\end{align}\\
\begin{figure}[H]
\hspace{-0.5cm}\includegraphics[scale=0.6]{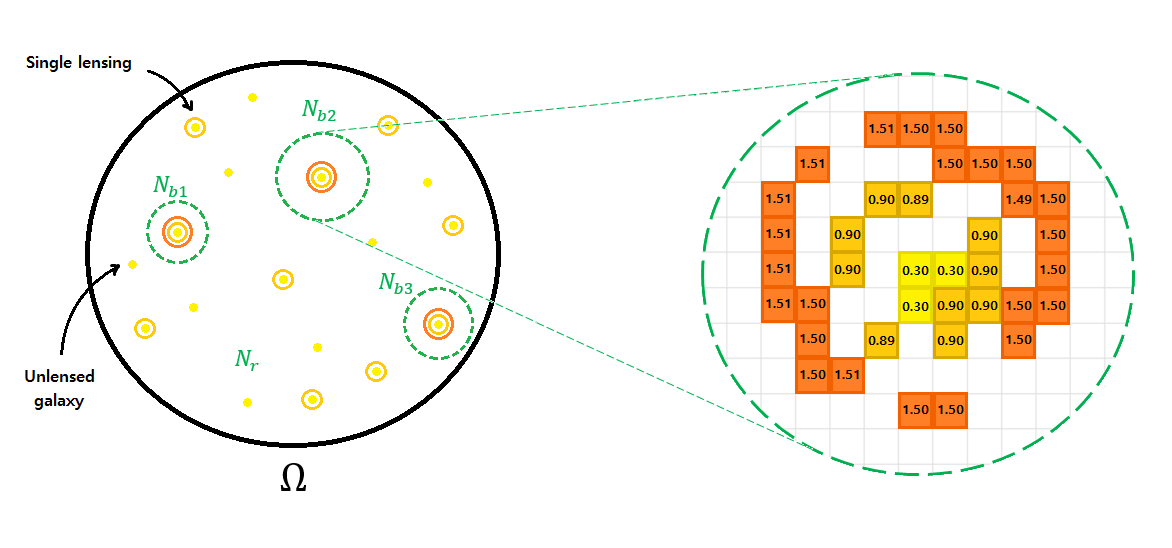}
\footnotesize
\caption{\small The left figure is a sky patch $\Omega$ that contains either photometric or spectroscopic pixel data. All the non-double lensing events are treated as noise. The inside area of the dotted regions (blobs) where the candidate systems are found are segmented from the sky and labeled as $N_{b1}, N_{b2}\cdots N_{bs}$. The rest the region is denoted as $N_r$, where its likelihood is expected to be significantly smaller. Region $N_{b2}$ is shown in detail on the right side to illustrate the pixel information. The numbers enclosed in each pixel are the redshift values.}\label{PatchskyFigure}
\end{figure}
Suppose we are given photometric or spectroscopic, pixel data of a patch $\Omega$ of the sky map (FIG \ref{PatchskyFigure}). We also assume every candidate (double lensed system) to be localized and non-overlapped with each other in the map. Two rings formed by two different lenses being overlapped with each other is geometrically impossible. The images would have been affected by each other's lens as well, thus, causing their ring shapes to be deformed. Rings must share a single lens to be overlapped while keeping their shapes, which is almost unlikely to happen. Therefore, it is reasonable to neglect the correlations between blobs. This can allows us to segment each of those interesting regions into a blob, and separate the integral part of the equation (\ref{BayesianFormalismSimplified}) into the blob region and the rest region. This is to effectively reduce expected costs for performing such statistics. If we were careful about selecting the blobs, the likelihood for the rest sky region to contain a double lensed system will be significantly smaller. Therefore, the equation (\ref{BayesianFormalismSimplified}) only needs to sum over the number of the candiate system. The local likelihood $P(data|N)$ contains gaussian functions for each individual local parameter. This term accounts for sets of any set of measurement errors. As a reminder once again, the local parameters were the three types of redshift and the angle ratio in this case. This means that the redshift parts are completely uncorrelated to each other.\\
\\
\begin{displaymath}\label{CovMatrix}
\Delta X^{-2}_{local}=\begin{bmatrix} \Delta X^{-2}_{local(1)} & 0 & 0 & 0 \\ 0 & \Delta X^{-2}_{local(2)} & 0 & 0 \\ 0 & 0 & \ddots & \vdots \\ 0 & 0 & \ldots & \Delta X^{-2}_{local(N_S)} \end{bmatrix}~~~~~,~~~~~\Delta X^{-2}_{local(i)}=\begin{bmatrix} \Delta z^{-2}_{L(i)} & 0 & 0 \\ 0 & \Delta z^{-2}_{S1(i)} & 0 \\ 0 & 0 & \Delta z^{-2}_{S2(i)} \end{bmatrix}
\end{displaymath}\\
\\
If these quantities are correlated, we should rather have a covariant Gaussian distribution with a covariant matrix accounting for the errors.\\
\begin{equation}
\setlength{\jot}{20pt}
\begin{aligned}\label{GaussianTerm}
&P(data|N_0,N_1,N_2 \cdots N_S)=~~\prod_{N_i=0}^{N_S}~~\frac{1}{(2\pi)^{N_S/2}\Delta X_{local(i)}}e^{-\dfrac{{\left(data- X_{local(i)}\right)^2}}{2\Delta X^2_{local(i)}}}\\
&=~~\prod_{N_i=0}^{N_S}~~\frac{1}{(2\pi)^{3N_S/2}\Delta z_{L(i)}\Delta z_{S1(i)}\Delta z_{S2(i)}}e^{-\dfrac{{(data[z_{L(i)}]-z'_{L(i)})^2}}{2\Delta z^2_{L(i)}}-\dfrac{{(data[z_{S1(i)}]-z'_{S1(i)})^2}}{2\Delta z^2_{S1(i)}}-\dfrac{{(data[z_{S2(i)}]-z'_{S2(i)})^2}}{2\Delta z^2_{S2(i)}}}\\
\end{aligned}
\end{equation}\\
where the template for no source does not exist (i.e. $X_{local(0)}=\emptyset$). Local prior $P(N|\Omega_M,\omega_{\Lambda})$ is a normalized version of the equation (\ref{DoublePD}).\\
\begin{align}\label{localLH}
P(data|N_S,\Omega_M,\omega_{\Lambda},f_{sky})=~~&\prod_{N_i=0}^{N_S}~~P(X_{local,i})\notag \\
=~~&\prod_{N_i=0}^{N_S}~~\frac{dN_i}{dz'_{L,i}dz'_{S1,i}dz'_{S2,i}}
\end{align}\\
The local prior is going to be different for each pair of $(\Omega_M, \omega_{\Lambda})$ which corresponds to the global prior. In order to save time for obtaining a posterior distribution, one could construct local prior templates in advance. This can be done by reconstructing the equation (\ref{DoublePD}) as functions of only local parameters (in this case the redshifts of the lens and sources) through fits. The smoothness of a posterior will highly depend on how well one can create the fitting templates. Whether to include $d\eta'_i$ in the denominator in the equation (\ref{localLH}) and another gaussian term in the equation (\ref{GaussianTerm}) depends on whether the parameter was considered for computing the expected number previously. If we have a perfect data (i.e. no false detections) the likelihood turns into a delta function $\delta_{Ns,\overline{N}}$, and the posterior will be left with just the Poisson term which is precisely the expression we had for the "Regular Bayesian Statistics-Expected Number".\\
\subsubsection*{Evidence ratio}
A surveyed data may contain a large number of detections that some analytical steps may cost expensively. For no detections, the equation (\ref{BayesianFormalismSimplified}) is 
\begin{align}\label{BayesianForZeroSystem}
P(\Omega_M,\omega_{\Lambda},(N=0)|data,f_{sky})\simeq~ e^{-(f_{sky}\overline{N})}
\end{align}\\
This is because of the fact that the local prior templates are used only for the cases when we do have a source or more.
For one detection, it is\\
\begin{align}\label{BayesianForOneSystem}
P(\Omega_M,\omega_{\Lambda},(N=1)|data,f_{sky})\simeq~ e^{-(f_{sky}\overline{N})}~+~\frac{(f_{sky}\overline{N})^{1}}{1!}e^{-(f_{sky}\overline{N})} \times I_1
\end{align}\\
where\\
\begin{align}\label{integralterm}
I_1=\int_{b1} P(data|N_1)P(N_1|\Omega_M,\omega_{\Lambda})~dX_{1}
\end{align}\\
For two detections, it is\\
\begin{align}\label{BayesianForTwoSystem}
P(\Omega_M,\omega_{\Lambda},(N=2)|data,f_{sky})\simeq~ e^{-(f_{sky}\overline{N})} ~+~ \frac{(f_{sky}\overline{N})^{1}}{1!}e^{-(f_{sky}\overline{N})}\times(I_1+I_2)
~+~\frac{(f_{sky}\overline{N})^{2}}{2!}e^{-(f_{sky}\overline{N})}\times 2!I_1I_2
\end{align}\\
where\\
\begin{align}\label{multipleintegralterm}
I_1I_2=\int_{b1}\int_{b2} P(data|N_1)P(data|N_2)P(N_1|\Omega_M,\omega_{\Lambda})P(N_2\Omega_M,\omega_{\Lambda})~dX_{1}dX_{2}
\end{align}\\
For three detections, it is\\
\begin{equation}
\setlength{\jot}{20pt}
\begin{aligned}\label{BayesianForThreeSystem}
P(\Omega_M,\omega_{\Lambda},(N=3)|data,f_{sky})\simeq~ &e^{-(f_{sky}\overline{N})} ~+~ \frac{(f_{sky}\overline{N})^{1}}{1!}e^{-(f_{sky}\overline{N})}\times(I_1+I_2+I_3)\\
~+~&\frac{(f_{sky}\overline{N})^{2}}{2!}e^{-(f_{sky}\overline{N})}\times 2!(I_1I_2+I_2I_3+I_1I_3)~+~\frac{(f_{sky}\overline{N})^{3}}{3!}e^{-(f_{sky}\overline{N})}\times 3!I_1I_2I_3\\
\end{aligned}
\end{equation}\\
Since we have to consider all possible combinations, the computation will become extremely time consuming as the number of detections increases. M.Johnson (2013) introduces 'evidence ratio technique' as the resolution for it. As previously discussed, the equation (\ref{BayesianFormalismSimplified}) can be evaluated separately for a blob region, $b$, and the rest region, $r$. Next, we want to separate the dataset into a data in the blob regions and a data in the rest region as well ($data=data_b+data_r$). 
\\
\begin{equation}
\setlength{\jot}{10pt}
\begin{aligned}\label{datasets}
data=&~[~data_r,~z_{L(1)},z_{S1(1)},z_{S2(1)}~,~z_{L(2)},z_{S1(2)},z_{S2(2)}~,~\cdots~~z_{L(N_S)},z_{S1(N_S)},z_{S2(N_S)}~]\\
X_{local,1}=&~[~~~~\emptyset~~~,~z'_{L(1)},z'_{S1(1)},z'_{S2(1)}~,~~~0~~~,~~~0~~~,~~~0~~~,~\cdots~~~~~0~~~~,~~~~0~~~~,~~~~0~~~~~]\\
X_{local,2}=&~[~~~~\emptyset~~~,~~~0~~~,~~~0~~~,~~~0~~~,~z'_{L(2)},z'_{S1(2)},z'_{S2(2)}~,~\cdots~~~~~0~~~~,~~~~0~~~~,~~~~0~~~~~]\\
&~~~~~~~~~~~~~~~~~~~~~~~~~~~~~~~~~~~~~~~~~~~~~\vdots\\
X_{local,N_S}=&~[~~~~\emptyset~~~,~~~0~~~,~~~0~~~,~~~0~~~,~~~0~~~,~~~0~~~,~~~0~~~,~\cdots ~~z'_{L(N_S)},z'_{S1(N_S)},z'_{S2(N_S)}~]\\
\end{aligned}
\end{equation}\\
\\
The dataset contains both the regions so we can distinguish it by expanding the Gaussian term in the likelihood expression\\
\begin{equation}
\setlength{\jot}{20pt}
\begin{aligned}\label{LHexpanded}
e^{-\dfrac{\left((data_b-X_{local})+(data_r-\emptyset)\right)^2}{2\Delta X_{local}^{2}}}=&~e^{-\dfrac{(data_b-X_{local})^2+(data_r)^2+2(data_b-X_{local})\cdot(data_r)}{2\Delta X^{2}_{local}}}\\
=&~e^{-\dfrac{(data_b-X_{local})^2+(data)^2-(data_b)^2-2(data_r)\cdot X_{local}}{2\Delta X^{2}_{local}}}\\
=&~e^{-\dfrac{(data_b-X_{local})^2+(data)^2-(data_b)^2}{2\Delta X^{2}_{local}}}\\
\end{aligned}
\end{equation}\\
where  $(data_r)\cdot X_{local}=0$. Therefore, after rearranging the equation (\ref{GaussianTerm}) we get\\
\begin{equation}
\setlength{\jot}{20pt}
\begin{aligned}\label{evidenceratioderivation}
P(data|N_0,N_1,N_2 \cdots N_S)=&~~\prod_{N_i=0}^{N_S}~~\frac{1}{(2\pi)^{N_S/2}\Delta X_{local(i)}}~e^{-\dfrac{(data)^2}{2\Delta X^{2}_{local,(i)}}}~\times~\frac{e^{-\dfrac{{\left(data_b- X_{local(i)}\right)^2}}{2\Delta X^2_{local(i)}}}}{e^{-\dfrac{(data_b)^2}{2\Delta X^{2}_{local(i)}}}}\\
=&~~P(data|N_0)~\times~~\prod_{N_i=1}^{N_S}~~\frac{e^{-\dfrac{{\left(data_b- X_{local(i)}\right)^2}}{2\Delta X^2_{local(i)}}}}{e^{-\dfrac{(data_b)^2}{2\Delta X^{2}_{local(i)}}}}\\
\end{aligned}
\end{equation}\\
The equation (\ref{BayesianForOneSystem}) can then be rewritten as \\
\begin{align}\label{evidenceratio}
P(\Omega_M,\omega_{\Lambda},(N=1)|data,f_{sky})\simeq~ e^{-(f_{sky}\overline{N})}~+~\frac{(f_{sky}\overline{N})^{1}}{1!}e^{-(f_{sky}\overline{N})}P(data|N_0)\times\rho_1
\end{align}\\
where\\
\begin{align}\label{ratioterm}
\rho_1=\int_{b_1} \frac{e^{-\dfrac{{\left(data_b- X_{local(1)}\right)^2}}{2\Delta X^2_{local(1)}}}}{e^{-\dfrac{(data_b)^2}{2\Delta X^{2}_{local(1)}}}}P(N_1|\Omega_M,\omega_{\Lambda})~dX_{1}
\end{align}\\
This is the evidence ratio that we may interpret as how 'useful' the template was for analyzing the data. In other words, if the template fit the data better than the template with a random field. In case it was 'useful', the evidence ratio will certainly become $\rho>1$. Conversely, $\rho\leq 1$ means the template provided made more mess than keeping the raw data as it is. Similarly, the equation (\ref{BayesianForTwoSystem}) is going to be\\
\begin{equation}
\setlength{\jot}{20pt}
\begin{aligned}\label{evidenceratio2}
P(\Omega_M,\omega_{\Lambda},(N=2)|data,f_{sky})\simeq&~ e^{-(f_{sky}\overline{N})} ~+~ \frac{(f_{sky}\overline{N})^{1}}{1!}e^{-(f_{sky}\overline{N})}P(data|N_0)\times(\rho_1+\rho_2)\\
~+&~\frac{(f_{sky}\overline{N})^{2}}{2!}e^{-(f_{sky}\overline{N})}P(data|N_0)\times 2!\rho_1\rho_2\\
\end{aligned}
\end{equation}\\
 The pattern seems to assure that the highest order term in the posterior will always be significantly greater than any other lower order terms. \\

\subsection{Hierarchical Bayesian Statistics - Expected numbers \& Angle ratios}
\begin{figure}[H]
\hspace{-1.5cm}\includegraphics[scale=0.23]{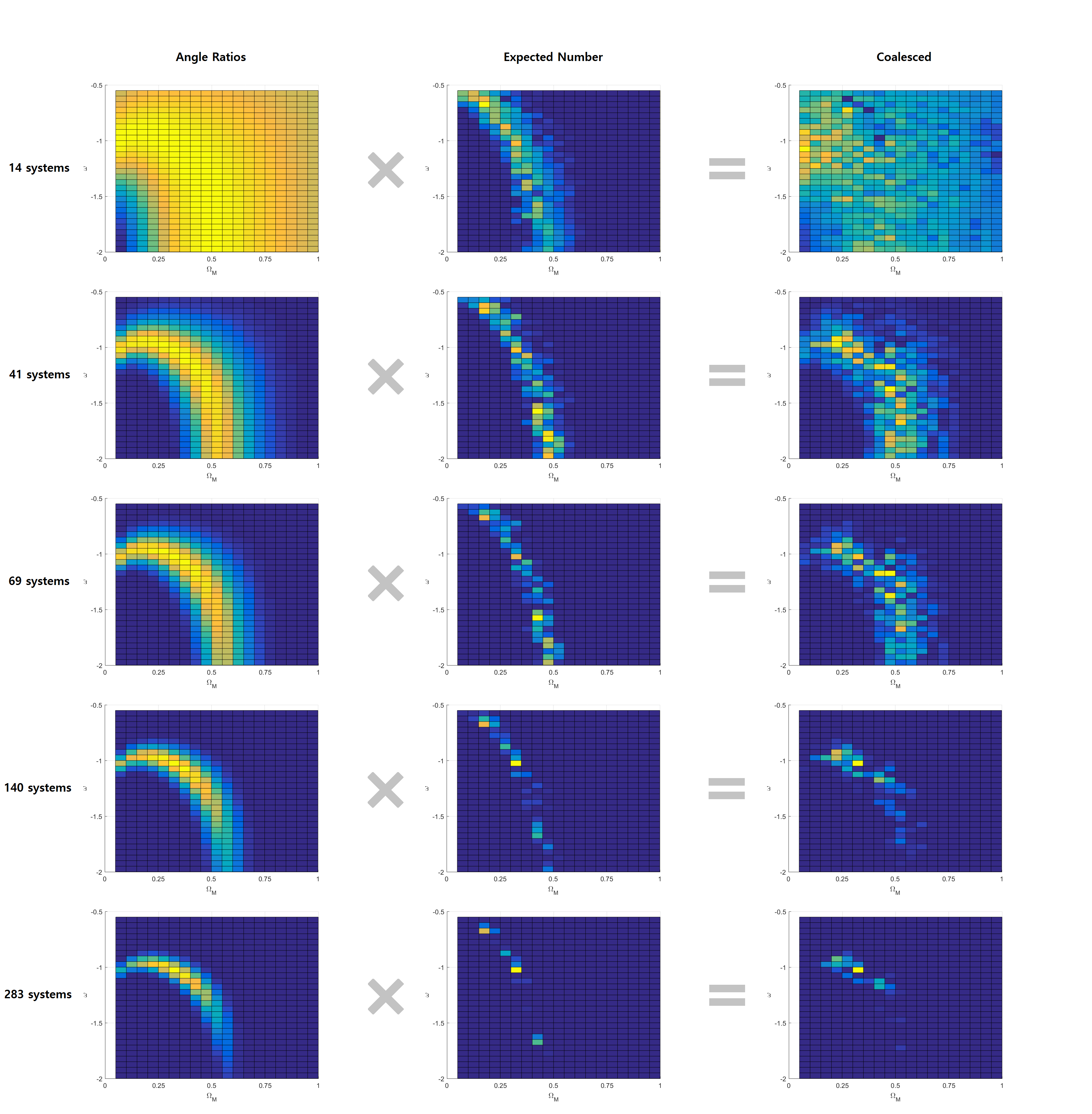}
\footnotesize
\caption{\small Hierarchical Bayesian Statistics (Top view).}\label{TopPlotsFigure}
\end{figure}

\begin{figure}[H]
\hspace{-1.5cm}\includegraphics[scale=0.23]{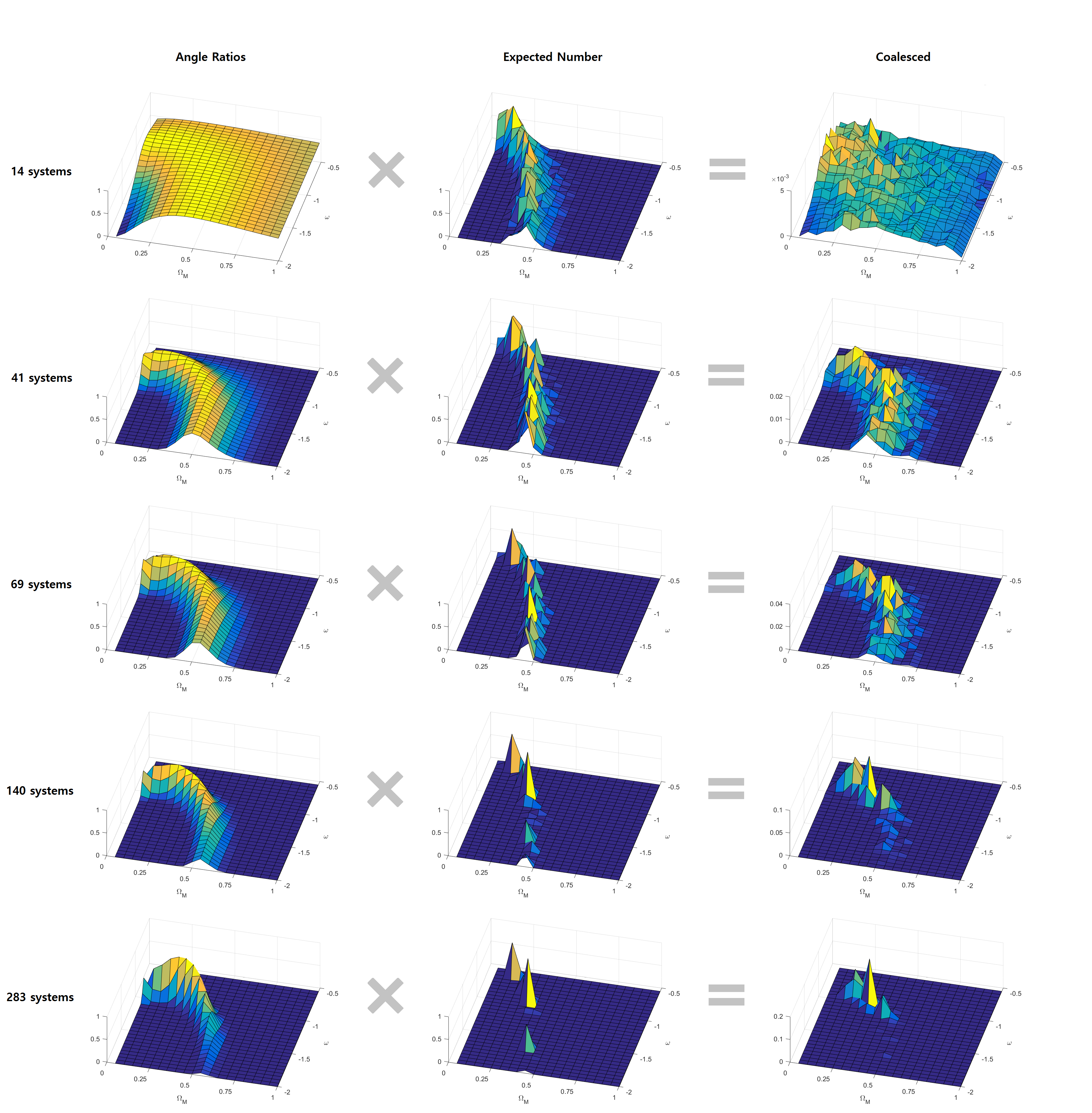}
\footnotesize
\caption{\small Hierarchical Bayesian Statistics (Side view).}\label{SidePlotsFigure}
\end{figure}

If a surveyed dataset also contains direclty measured angle ratios, we can combine the Regular Bayesian Statistics (angle ratio) previously introduced with the Hierarchical Bayesian Statistics (expected numbers) to further constrain the comsological parameters. Although the angle ratio will be an additional measured quantity, the parameter corresponding to the angle ratio can simply be a combination of the three variables that already exist for the redshifts (i.e. $\eta'=\eta'(z_L,z_{S1},z_{S2})$). Hence, a gaussian term for the angle ratio will appear in the local likelihood while its corresponding local prior will be absent assuming we do not need any cut-off function for a minimum resolution.
\\
\begin{align}\label{CoalescedBayesian}
P(\Omega_M,\omega_{\Lambda},|data,f_{sky})~\simeq~(f_{sky}\overline{N})^{N_S}e^{-(f_{sky}\overline{N})}P(data|N_0)\times\prod_{N_i=1}^{N_S}\sigma_i\\
\end{align}\\
where\\
\begin{align}\label{Coalescedratio}
\sigma_i=\int_{b_i} \frac{e^{-\dfrac{{\left(data_b- X_{local(i)}\right)^2}}{2\Delta X^2_{local(i)}}}}{e^{-\dfrac{(data_b)^2}{2\Delta X^{2}_{local(i)}}}}P(N_1|\Omega_M,\omega_{\Lambda})~e^{-\dfrac{\left(data\left[\eta_i\right]-\eta'_{(i)}\right)^2}{2\Delta\eta^2_{(i)}}}~dX_{i}
\end{align}\\
~
FIG \ref{TopPlotsFigure} and FIG \ref{SidePlotsFigure} show the results from the Hierarchical Bayesian Statistics for the expected numbers and the angle ratios with $\Delta z=0.005$ and $\Delta \eta=0.01$. In this case we did not account for the angle ratio for calculating the expected numbers, so it was perfectly acceptable to separately evaluate the overall posterior (the posterior from the angle ratios and the posterior from the expected numbers, combined). With a large number of detections, the results start to clearly indicate where the parameters that we are seeking are located. The projected degeneracies (from the way the probability densities are distributed) were slightly different between the results of the two statistics, which they can further constrain the parameters by having a small intersecting region. This implies that combining the two statistics is a very favorable idea.
\subsection{Conclusion}
We have set up the toy model for generating the population of double-lensing events, which involved several steps. First step was defining the strong-lensing event. There exist many other ways of defining the lensing event, and in this paper we have claimed that a lensed image arc with its size of greater than $\pi/2$ is the strong lensing event. We assumed our lens and sources to be the SIS objects for the simplification. From these assumptions, we could obtain the probability cone for a single lensing event. Next, by applying the Schechter's luminosity function and stricting the lower boundary based on the lensing magnification bias we could roughly estimate the number densities of objects as a function of the comoving distance between the event and the observer. Expected number of events was then obtained by multiplying the probability of double-lensing events at a given field of view. The result showed that our expected number was far beyond the number of detected lensing events from the Hubble data \cite{key1}. As estimations made by P. Marshall (2005) and G. Dobler (2008) also seem to agree with our prediction, we doubt that oversimplification may have been the only cause for such consequence, and believe that a lot more lensing events should be discovered in the future. The expected number of the double-lensing events was also computed, which we concluded that the number is high enough to perform statistical analysis. Assuming we made a good approximation on the lensing population, we have introduced that properties of dark energy can be measured in the Hierarchical Bayesian scheme. To explain the method we first arbritrariily defined the cosmological parameters and computed their corresponding expected number of the events. Then, we pretended that the expected number are the actual number of events observed and the parameters are unknown. We successfully showed that one can retrieve the parameters by setting a prior distribution and its likelihood then obtaining the posterior distribution for the unknown parameters. By including the double-ring ratio which only depends on the density parameters as another observable quantity, the Hierarchical Bayesian method became even more useful to probe the dark energy. Since we have ignored the multiformity of galaxies, more generality should be allowed in constructing the toy model. 
\newpage

~
~
~

\newpage
~
\centerline{\Large Appendix A : Hierarchical Bayesian Statistics}
\\
\\
\\
~

We first build the local prior templates for each of the global priors. A perfect data is directly input in all the templates to yield a posterior distribution. If data is not perfect, we add a Gaussian likelihood to the local priors before proceeding the calculations. The template (a single grid in the posterior) that matches the most with the data will show the highest peak in the posterior distribution.\\
\begin{figure}[H]
\hspace{-1.5cm}\includegraphics[scale=0.3]{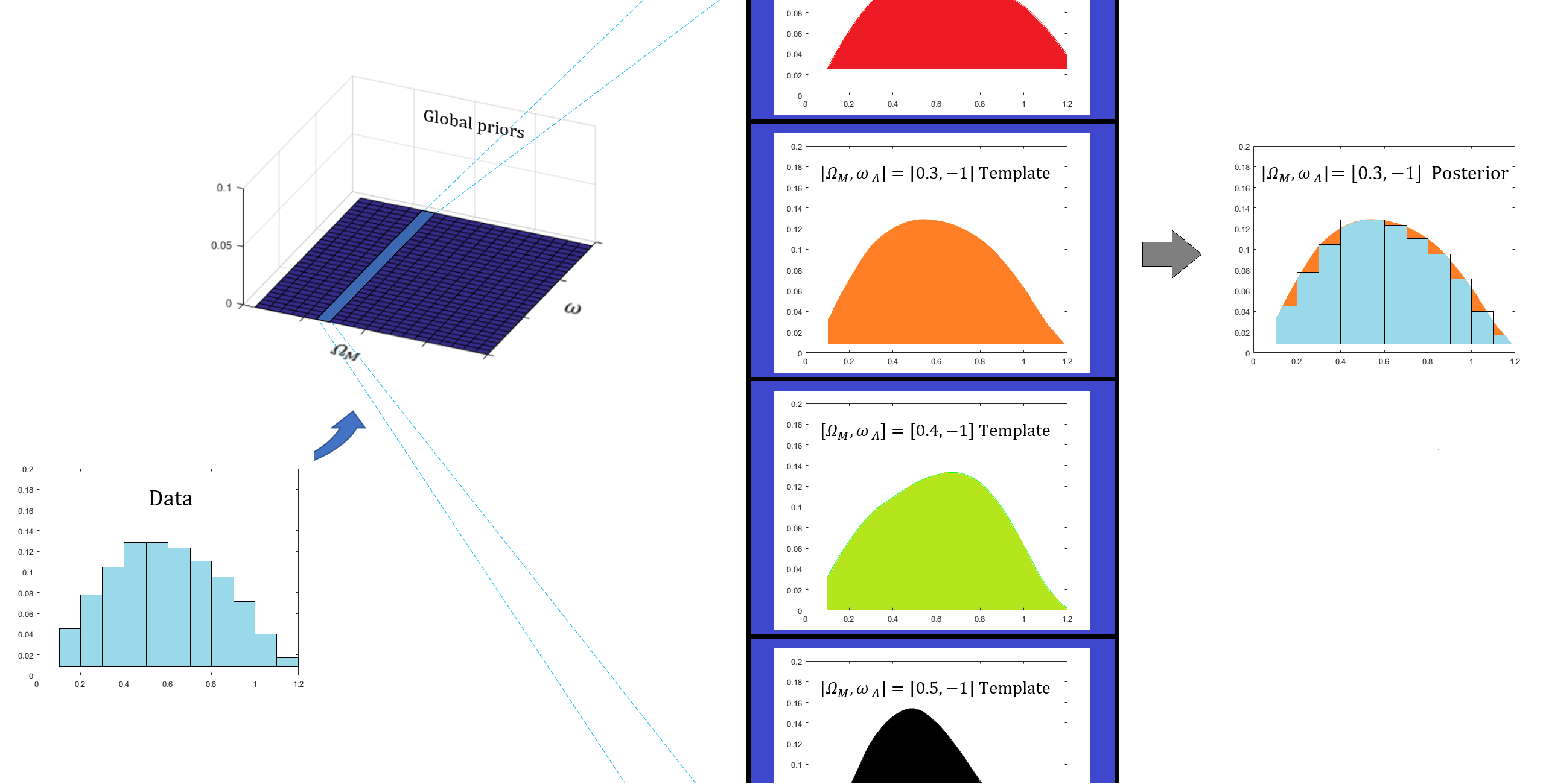}
\footnotesize
\caption{\small Bayesian Process.}\label{BayesianProcessFigure}
\end{figure}
Poor fitting for the templates may not result in a smooth posterior distribution, and FIG (\ref{PoissonOnlyFigure}) is an example for such a case.
\end{document}